\documentclass[twocolumn,useAMS,usenatbib]{mn2e}
\usepackage{graphicx,amsmath}
\usepackage{bm}

\newcommand{\rmd}{{\rm d}}
\newcommand{\rme}{{\rm e}}
\newcommand{\rmi}{{\rm i}}

\newcommand{\bv}{{\bmath v}}
\newcommand{\bx}{{\bmath x}}
\newcommand{\bk}{{\bmath k}}
\newcommand{\dm}{\delta_{\rm m}}
\newcommand{\dg}{\delta_{\rm g}}
\newcommand{\tdm}{\tilde\delta_{\rm m}}
\newcommand{\tdg}{\tilde\delta_{\rm g}}
\newcommand{\Pm}{P_{\rm m}}
\newcommand{\Pme}{P_{{\rm m}\epsilon}}
\newcommand{\Pg}{P_{\rm g}}
\newcommand{\nhat}{\hat{\bmath n}}
\newcommand{\Lhat}{\hat{\bmath L}}
\newcommand{\ehat}{\hat{\bmath e}}
\newcommand{\mR}{{\mathbfss R}}
\newcommand{\bs}{{\bmath s}}
\newcommand{\nseff}{n_{\rm s}^{\rm eff}}

\title[Tidal alignments]
{Tidal alignments as a contaminant of redshift space distortions}

\author[Hirata]
 {Christopher M. Hirata
\\Caltech M/C 350-17, Pasadena, California 91125, USA}

\date{24 June 2009}

\begin{document}
\maketitle

\begin{abstract}
We investigate the effect of orientation-dependent selection effects on galaxy clustering in redshift space.  It is found that if galaxies are aligned 
by large-scale tidal fields, then these selection effects give rise to a dependence of the observed galaxy density on the local tidal field, in 
addition to the well-known dependences on the matter density and radial velocity gradient.  This alters the galaxy power spectrum in a way that is 
different for Fourier modes parallel to and perpendicular to the line of sight.  These tidal galaxy alignments can thus mimic redshift space 
distortions, and thus result in a bias in the measurement of the velocity power spectrum.  If galaxy orientations are affected only by the local tidal 
field, then the tidal alignment effect has exactly the same scale and angular dependence as the redshift space distortions in the linear regime, so it 
cannot be projected out or removed by masking small scales in the analysis.  We consider several toy models of tidal alignments and 
orientation-dependent selection, normalize their free parameter (an amplitude) to recent observations, and find that they could bias the velocity 
amplitude $f(z)G(z)$ by 5--10 per cent in some models, although most models give much smaller contamination.  We conclude that tidal alignments may be 
a significant systematic error in redshift space distortion measurements that aim to test general relativity via the growth of large-scale structure.  
We briefly discuss possible mitigation strategies.
\end{abstract}


\section{Introduction}

Redshift space distortions (RSDs) have a long history as a cosmological probe.  \citet{1987MNRAS.227....1K} first showed that even on very large 
scales, the galaxy power spectrum in redshift space is significantly affected by bulk flows, and proposed this as a technique to measure $\Omega_{\rm 
m}$.  The anisotropy of the correlation function was explored by \citet{1992ApJ...385L...5H}.  The redshift-space distorition technique was first 
applied to the {\slshape Infrared Astronomical Satellite} ({\slshape IRAS}) 2 Jy redshift survey by \citet{1993ApJ...406L..47H}, yielding a measurement 
of $\Omega_{\rm m}=0.5^{+0.5}_{-0.25}$.  Subsequent work with {\slshape IRAS}-selected surveys \citep{1995MNRAS.275..515C} and optically-selected 
redshift surveys \citep{1996ApJ...468....1L, 1998MNRAS.296..191R} showed that the redshift-space anisotropy parameter $\beta=\Omega_{\rm m}^{0.6}/b$ 
was $\sim 0.5$, providing strong evidence that either $\Omega_{\rm m}<1$ or a high bias $b\sim 2$ was required.  The 2-degree Field (2dF) survey 
represented a large step forward for redshift space distortion measurements, with \citet{2001Natur.410..169P} reporting $\beta=0.43\pm0.07$.  
Combining this approach with the galaxy bispectrum, which independently gives $b$, allowed \citet{2002MNRAS.335..432V} to measure $\Omega_{\rm 
m}=0.27\pm 0.06$, completely internal to the 2dF survey.  The Sloan Digital Sky Survey (SDSS) has reported RSD measurements from both 
magnitude-limited and color-selected (luminous red galaxy; LRG) samples \citep{2004ApJ...606..702T, 2006PhRvD..74l3507T}. RSD measurements have also 
been reported for quasars \citep{2005MNRAS.360.1040D} and more recently (and with higher signal-to-noise ratio) for galaxies at $z>0.5$ 
\citep{2007MNRAS.381..573R, 2008Natur.451..541G}.

\begin{figure*}
\includegraphics[width=5.8in]{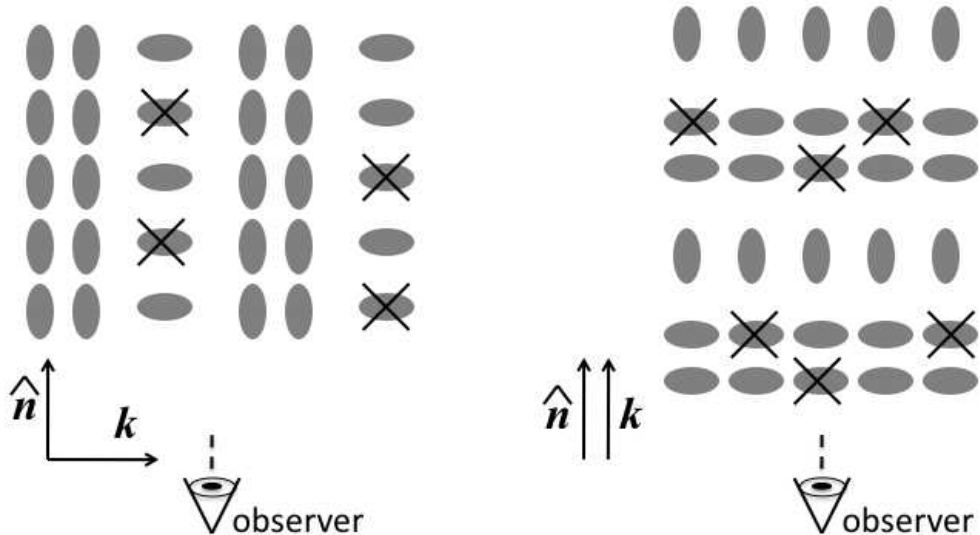}
\caption{\label{fig:cartoon}The anisotropic selection effect.  We show a Fourier mode $\bk$ of the density field, oriented transverse to the line of 
sight in the left panel and radially in the right panel.  The galaxies are shown aligned along the stretching direction of the tidal field, as 
appropriate for LRGs (but exaggerated).  If selection effects prefer galaxies where the observer's line of sight looks down the long axis, then some 
galaxies whose short axes point toward the observer are lost (marked with an X in the figure).  The remaining galaxies show stronger clustering for 
transverse modes ($\bk\perp\nhat$) than for radial modes ($\bk\parallel\nhat$), thereby contamination redshift-space distortion measurements.  In this 
case the effect is opposite to the linear Kaiser effect and hence biases estimates of the velocity power spectrum low.}
\end{figure*}

Today, other methods of constraining $\Omega_{\rm m}$ are more powerful, namely combining the cosmic microwave background (CMB) with supernovae and 
baryon acoustic oscillation (BAO) measurements \citep{2009ApJS..180..306D}.  RSDs have also developed a reputation for having difficult-to-control 
systematics since nonlinear evolution affects them at larger scales than the real space power spectrum.  However, the discovery of dark energy has 
driven a revival in the subject of redshift space distortions.  The same jumbo redshift surveys that are required by the BAO method will measure many 
modes of the galaxy field at large scales -- precisely what is required to exploit the power of the redshift-space distortion technique in the regime 
where its systematics should be minimized \citep{2008arXiv0810.1518W}.  Modified gravity theories in which the acceleration of the universe reflects a 
breakdown of general relativity rather than a new contribution to the cosmic energy budget have motivated studies of the growth of structure, as it is 
impossible to distinguish these models from dark energy using distance information alone (e.g. \citealt{2006PhRvD..74d3513I, 2008arXiv0807.0810S, 
2009JCAP...01..048S}).  In this case redshift-space distortions generally measure the velocity power spectrum $P_v(k)$.  Finally, advances in 
understanding nonlinear evolution with larger $N$-body simulations and with halo occupation models \citep{2002PhR...372....1C}, and novel techniques 
using multiple galaxy tracers to reduce the uncertainty on $P_v(k)$ \citep{2008arXiv0810.0323M} have made the case for RSDs more compelling.  Indeed, 
RSDs are part of the prime science case of the proposed {\slshape Euclid} dark energy mission.

RSDs suffer from a very different set of limitations than alternative probes of the growth of cosmic structure such as cosmic shear (CS), the 
integrated Sachs-Wolfe (ISW) effect, and the cluster mass function.  CS directly probes the matter distribution without assumptions about galaxy 
biasing; it is however technically difficult to measure galaxy shapes with sufficient control of systematics, and practical implementations of CS rely 
on photometric redshifts.  The ISW effect has turned out to be relatively clean \citep{2008PhRvD..77l3520G, 2008PhRvD..78d3519H}, but its statistical 
power suffers from severe cosmic variance limitations.  Clusters are easier to observe than CS, and are numerous so that statistical errors can be made 
small even with present data; the main uncertainty is instead in the astrophysics of relating cluster observables to virial masses.

The most worrying systematic error in the RSD method is the behaviour of galaxy biasing and velocities in the nonlinear regime.  However, we show in 
this paper that the alignment of galaxies by large-scale tidal fields results in an additional systematic error.  If galaxies are preferentially 
aligned along the stretching axis of the tidal field, as is the case for LRGs, and there is any viewing direction-dependent selection effect (e.g. preferring 
galaxies where we look down the long axis, as would occur with an aperture magnitude cut), then Fourier modes of the density field along the line of sight are suppressed because the galaxies in the 
troughs are more likely to be selected than galaxies in the crests.  This is shown in cartoon form in Fig.~\ref{fig:cartoon}.  Tidal alignments of disk galaxies would if present produce a similar effect 
because these galaxies are often selected by optical continuum or emission line luminosity cuts, which suffer inclination-dependent internal extinction.
The alignment effect is small -- we will argue that in pessimistic scenarios it 
contaminates redshift space distortion measurements by 5--10 per cent -- but we will show that it exactly mimics the angular and scale dependence of RSDs, making it 
hard to remove.  Moreover, it is not included in mock catalogues that paint galaxies onto $N$-body simulations.

This paper is organized as follows.  In Section~\ref{sec:zspace}, we outline the standard theory of the RSD technique.  In Section~\ref{sec:problem}, 
we show how they can be altered by galaxies with tidal alignments and viewing direction-dependent selection effects.  Section~\ref{sec:t} specializes to the 
physically motivated case of galaxies aligned by the large-scale tidal field.  In Section~\ref{sec:micro} we consider some crude models of 
orientation-dependent selection effects and find that they may be significant.  Section~\ref{sec:cosmic} compares the effects of tidal alignments on 
RSD to their more familiar effect on CS.  We conclude and discuss mitigation strategies in Section~\ref{sec:disc}.

The early parts of this paper, especially Section~\ref{sec:problem}, are intended to be general, exploring all models allowed by symmetry; they are unavoidably heavy on mathematical formalism.  
Section~\ref{sec:micro}, in contrast, is intended to consider specific examples; it uses rough calculations since orientation-dependent selection effects are not amenable to precise calculation.  
Section~\ref{sec:cosmic} is included mainly to develop intuition and hence also makes use of some rough calculations.

\section{Standard redshift space distortions}
\label{sec:zspace}

We begin by reviewing the standard theory of redshift-space distortions \citep{1987MNRAS.227....1K}. In this paper, we denote the galaxy bias by $b$ 
and the rate of growth of structure by $f=\rmd\ln G/\rmd\ln a$, where $G$ is the growth function.

We consider an underlying matter density fluctuation field $\dm(\bx)$, where $\bx=(x_1,x_2,x_3)$ is the comoving position.  We work in the flat-sky 
approximation and take the observer's line of sight to be in the $x_3$-direction.  The bulk velocity field is described in linear perturbation theory 
by
\begin{equation}
\bv(\bx) = -f\nabla_\bx \nabla_\bx^{-2} \dm(\bx).
\label{eq:v}
\end{equation}
For a sample of galaxies with linear bias $b$, the real-space density fluctuation of galaxies in the linear regime is $\dg^{\rm(r)}(\bx)=b\dm(\bx)$.  
However, in redshift space, the density is also corrected by the Jacobian of the conversion from real to redshift space, $1-\partial v_3/\partial 
x_3$.\footnote{In linear theory we may neglect the change in the position ${\bx}$ as we move from real to redshift space.}
That is,
\begin{equation}
\dg(\bx) = b\dm(\bx) - \frac{\partial v_3(\bx)}{\partial x_3}.
\label{eq:dg}
\end{equation}
These equations are most easily represented in Fourier space:
\begin{equation}
\tilde \bv(\bk) = -\rmi f \frac{\bk}{k^2}\tdm(\bx)
\end{equation}
and
\begin{equation}
\tdg(\bk) = b\tdm(\bk) - \rmi k_3\tilde v_3(\bk).
\end{equation}

The overall galaxy density in redshift space is then
\begin{equation}
\tdg(\bk) = (b + f\mu^2)\tdm(\bk),
\label{eq:tdgx}
\end{equation}
where $\mu=k_3/k$ is the cosine of the angle between the line of sight and the direction of the Fourier mode under consideration.  This implies an observed galaxy power spectrum,
\begin{equation}
\Pg(\bk) = (b + f\mu^2)^2\Pm(k).
\label{eq:Pg}
\end{equation}
The amount of anisotropy depends on the parameter $\beta\equiv f/b$.

The standard use of redshift space distortions is to use the amplitudes $\Delta_{\rm g}^2(\bk) = k^3\Pg(\bk)/2\pi^2$, which satisfy
\begin{equation}
\Delta_{\rm g}(\bk) = b\Delta_{\rm m}(k) + \mu^2f\Delta_{\rm m}(k),
\label{eq:dgk}
\end{equation}
to extract the quantity $f\Delta_{\rm m}(k)$.  If the matter power spectrum is normalized at high redshift by the CMB, then $\Delta_{\rm m}(k)\propto G$, and redshift space distortions can be used to obtain the quantity $f(z)G(z)$.  Any contamination to Eq.~(\ref{eq:dgk}) produces a fractional error in $f(z)G(z)$ equal to the fractional change in the $\mu^2$ term, to the change in the coefficient of $\mu^2$ divided by $f\Delta_{\rm m}(k)$.

The above analysis assumes that when we measure a Fourier mode of the galaxy density field we know the calibration of the radial and transverse 
distance scales so that the comoving ${\bmath k}$ is known.  This could be done using the BAO combined with a ruler length set by the CMB, or in a 
parameterized cosmology by combining all cosmological probes.  The parameter $\beta$ only requires the relative calibration of $k_\parallel$ and 
$k_\perp$, which amounts to determining the product $H(z)D(z)$ where $H(z)$ is the Hubble rate and $D(z)$ is the comoving angular diameter distance.  
In principle this could be done internally to the galaxies using the \citet{1979Natur.281..358A} test.  In this case, extracting $f$ would require 
independent information on the galaxy bias, e.g. from the three-point function.

\section{The problem}
\label{sec:problem}

\subsection{Intrinsic alignments}
\label{ss:ia}

If galaxies were randomly oriented, Eq.~(\ref{eq:Pg}) would fully describe the linear regime.  However, some classes of real galaxies, including LRGs, exhibit 
correlations of their orientation with large-scale structure (e.g. \citealt{1982A&A...107..338B}; see \citealt{2007MNRAS.381.1197H} for a recent measurement of the 
correlation function).  This section defines the notation needed to handle selection effects that depend on galaxy orientation and viewing direction.

The galaxy orientation is described by 3 Euler angles $(\theta,\phi,\psi)$.  These define a rotation matrix ${\mathbfss Q}(\theta,\phi,\psi)\in$SO(3) that transforms 
the ``lab'' frame coordinates to a frame aligned with the galaxy; see 
Appendix~\ref{app:Euler} for explicit expressions.  The well-known volume element of SO(3) is
\begin{equation}
\rmd^3{\mathbfss Q} = \sin\theta\,\rmd\theta\,\rmd\phi\,\rmd\psi,
\end{equation}
and its total volume is $\int_{\rm SO(3)}\rmd^3{\mathbfss Q}=8\pi^2$.
We consider the conditional probability distribution ${\cal P}({\mathbfss Q}|\bx)$ for the orientation ${\mathbfss Q}$ of a galaxy at position $\bx$.  
In the case of randomly oriented galaxies we have ${\cal P}({\mathbfss Q}|\bx)=1/(8\pi^2)$, but in general this probability distribution may depend on the 
local environment (tidal field, density, etc.) at $\bx$.

Our next ingredient is the dependence of the observational selection function on the viewing geometry and galaxy orientation.  We assume that the observer is looking 
along the line of sight\footnote{We define this so that the observer points their telescope in direction $\nhat$; the light from
the galaxy is propagating in direction $-\nhat$.} $\nhat$.  The appearance of the galaxy then depends on the observer's line-of-sight expressed in the galaxy frame, 
i.e. $\hat{\bmath m}={\mathbfss Q}\nhat$.  It is therefore possible for the probability of selecting a galaxy to vary depending on ${\mathbfss Q}\nhat$:
\begin{equation}
P\propto 1 + \Upsilon({\mathbfss Q}\nhat,\bx).
\label{eq:PU}
\end{equation}
We may take the anisotropy of the selection function to have mean zero when averaged over all possible viewing directions,
\begin{equation}
\int_{S^2} \Upsilon(\hat{\bmath m},\bx)\,\rmd^2\hat{\bmath m} = 0,
\label{eq:mhat}
\end{equation}
since any ${\hat{\bmath m}}$-independent constant added to $\Upsilon$ can be absorbed into the local galaxy density.  Note that Eq.~(\ref{eq:mhat}) implies that the 
selection probability variation averaged over all possible orientations is also zero, i.e.
\begin{equation}
\int_{\rm SO(3)} \Upsilon({\mathbfss Q}\nhat,\bx)\,\rmd^3{\mathbfss Q} = 0.
\label{eq:so3int}
\end{equation}

The tidal alignment effect on large scale structure observations depends on the combination of the intrinsic alignment model, which determines ${\cal P}({\mathbfss 
Q}|\bx)$, and the dependence $\Upsilon$ of the selection function on geometry.  The number density of selected galaxies $N({\rm selected})$ is then related to 
the true number density of galaxies $N({\rm true})$ by averaging Eq.~(\ref{eq:PU}) over the distribution of galaxy orientations at $\bx$:
\begin{eqnarray}
\frac{N({\rm selected})}{N({\rm true})} &\propto& \int_{\rm SO(3)} {\cal P}({\mathbfss Q}|\bx) [1 + \Upsilon({\mathbfss Q}\nhat,\bx)] \rmd^3{\mathbfss Q}
\nonumber \\
&=& 1 + \int_{\rm SO(3)} {\cal P}({\mathbfss Q}|\bx) \Upsilon({\mathbfss Q}\nhat,\bx)\,\rmd^3{\mathbfss Q}.
\end{eqnarray}
This motivates defining
\begin{equation}
\epsilon(\nhat|\bx) = \int_{\rm SO(3)} {\cal P}({\mathbfss Q}|\bx) \Upsilon({\mathbfss Q}\nhat,\bx)\,\rmd^3{\mathbfss Q}.
\label{eq:epsilon}
\end{equation}
which represents the viewing direction dependent selection function at position $\bx$.
Examining Eq.~(\ref{eq:so3int}), we see that $\epsilon$ 
vanishes if either the galaxy orientations are isotropically distributed [in which case ${\cal P}({\mathbfss Q}|{\mathbf x})=1/(8\pi^2)$], or if $\Upsilon$ vanishes.  
Both intrinsic alignments (${\cal P}$ depends on ${\mathbfss Q}$) and orientation-dependent selection ($\Upsilon\neq 0$) are required to produce an effect on large 
scale structure.

[In the case of an axisymmetric disk galaxy, the third Euler angle $\psi$ is irrelevant.  In this case, integrals such as Eq.~(\ref{eq:epsilon}) reduce to integrals 
over the disk normal vector $\hat{\bmath L}$.]

The real-space density of observed galaxies at position $\bx$ as 
seen by an observer looking in direction $\nhat$ has fluctuation
\begin{equation}
1+ \dg^{\rm(r,obs)}(\bx,\nhat) = [1 + \dg^{\rm(r)}(\bx)][1+\epsilon(\nhat|\bx)].
\label{eq:dgro}
\end{equation}
In a real galaxy survey, the actual viewing direction $\nhat$ is $\ehat_3$, and our final computation of the galaxy properties will be obtained by substituting 
$\nhat=\ehat_3$ into Eq.~(\ref{eq:dgro}).

Equation~(\ref{eq:mhat}) implies that $\epsilon(\nhat|\bx)$ to average to zero over the unit sphere (i.e. over all possible viewing directions),
\begin{equation}
\int_{S^2} \epsilon(\nhat|\bx) \rmd^2\nhat = 0.
\label{eq:0}
\end{equation}

We then find, instead of Eq.~(\ref{eq:tdgx}), the observed redshift-space galaxy density in the linear regime is
\begin{equation}
\tdg(\bk) = (b+f\mu^2)\tdm(\bk) + \tilde\epsilon(\ehat_3|\bk).
\label{eq:tdgx2}
\end{equation}

\subsection{Statistics of intrinsic alignments}
\label{ss:stat}

In order to construct the observed galaxy power spectrum, we need to develop the statistics of $\epsilon(\nhat|\bx)$.  This is a random field as a function of a 
position ${\bmath x}$ and direction $\hat{\bmath n}$, and its 
statistical description can be achieved in the same way that one decomposes the CMB temperature $\delta T(\bx,\nhat)$ in cosmological perturbation theory 
\citep{1995ApJ...455....7M, 1997PhRvD..56..596H}.
One first Fourier transforms the spatial variables to get 
$\tilde\epsilon(\nhat|\bk)$.  Then one introduces a rotation matrix $\mR\in$SO$(3)$ that rotates the unit vector $\hat\bk$ to the $\ehat_3$ axis, i.e. $\mR\hat\bk=\ehat_3$.\footnote{Note that $\mR$ is 
not unique, since it amounts to an arbitrary choice of which unit vector in the plane orthogonal to $\bk$ gets rotated to $\ehat_1$.}  Here $\mR$ 
depends on $\hat\bk$, but we will not write this dependence 
explicitly to avoid confusion.  Finally, we decompose the dependence of $\tilde\epsilon(\nhat|\bk)$ on the viewing direction $\nhat$ into spherical harmonics,
\begin{equation}
\tilde\epsilon(\nhat|\bk) = \sum_{l=1}^\infty \sum_{m=-l}^l (-\rmi)^l \sqrt{\frac{4\pi}{2l+1}}\, \tilde\epsilon_{lm}(\bk) Y_{lm}(\mR\nhat).
\label{eq:elm}
\end{equation}
This corresponds to choosing the ``North Pole'' of the spherical harmonic basis to be in the direction $\hat\bk$.
We note that Eq.~(\ref{eq:0}) eliminates the $l=0$ term.

Translation invariance forces different $\bk$-modes to be independent, and rotational invariance around $\bk$ forces different different values of $m$ to be independent.  Therefore, we can write the 
power spectrum of $\epsilon$:
\begin{equation}
\langle \tilde\epsilon_{lm}^\ast(\bk) \tilde\epsilon_{l'm'}(\bk') \rangle = (2\pi)^3
P_\epsilon^{ll'm}(k) \delta_{mm'} \delta^{(3)}(\bk-\bk').
\label{eq:Pe}
\end{equation}
These power spectra are not arbitrary.  Appendix~\ref{app:real} shows that since $\epsilon(\nhat|\bx)$ is a real field,
$P_\epsilon^{ll'm}(k)$ must also be real.
Invariance under reflection across a plane containing $\bk$ forces the restriction
\begin{equation}
P_\epsilon^{ll'm}(k) = P_\epsilon^{ll',-m}(k),
\end{equation}
and by swapping the $l$ and $l'$ labels in Eq.~(\ref{eq:Pe}) and recalling that the power spectra are real, we find the symmetry relation:
\begin{equation}
P_\epsilon^{ll'm}(k) = P_\epsilon^{l'lm}(k).
\end{equation}
Therefore the power spectra are completely described by the cases with $0\le m\le l'\le l$, with $l,l'>0$.

In addition to its intrinsic power spectrum, $\epsilon$ can have a cross-power spectrum with the matter field.  This can be defined in analogy to Eq.~(\ref{eq:Pe}).  The matter field has the same symmetry as the $l=m=0$ mode of $\epsilon$ (if it existed), so by rotational symmetry only the $m=0$ components of $\epsilon$ can correlate with the matter field.  We thus have
\begin{equation}
\langle \tdm^\ast(\bk) \tilde\epsilon_{l0}(\bk') \rangle = (2\pi)^3
\Pme^{l}(k) \delta^{(3)}(\bk-\bk')
\label{eq:Pme}
\end{equation}
for $l>0$.  Again, $\Pme^l(k)$ must be real.

A further simplification is possible for galaxies that are inversion-symmetric, i.e. have the same appearance if viewed from the opposite direction.  This is true for 
optically thin, triaxial elliptical galaxies, and is true for disk galaxy models that are axisymmetric and have a reflection symmetry across their equatorial 
plane.  It may also be true in a {\em statistical} sense for real disk galaxies: while the specific arrangement of dust and H~{\sc ii} regions will not have 
inversion symmetry, it is plausible that these local imperfections would be unique to each galaxy and not correlated with large-scale structure.  In this case, 
$\epsilon(\nhat|\bx)=\epsilon(-\nhat|\bx)$ and hence $\tilde\epsilon_{lm}(\bk)=0$ for odd $l$.  All power spectra with odd $l$ or $l'$ then vanish.

In traditional cosmology language, the $m=0$ contributions to this equation are scalars, $m=\pm 1$ are vectors, $m=\pm 2$ are tensors, and so on.  One would expect in 
the linear regime that only the scalars ($m=0$) are present since large-scale density perturbations are scalars and cannot source any alignment of galaxies with 
$m\neq0$ symmetry.  (Physically, in the absence of mode coupling, a Fourier mode $\bk$ possesses rotational symmetry around the axis $\hat\bk$ that prevents any 
$m\neq0$ component.)
At small scales it is possible that the $m\neq 0$ alignments arise from nonlinear mode coupling.  Since Eq.~(\ref{eq:tdgx2}) is valid only when 
the density and velocity fields are in the linear regime, in the body of this paper we will only consider the scalar ($m=0$) intrinsic alignments.  We note, however, 
that the scale at which nonlinear effects start to influence intrinsic alignments is unknown and may be larger than the traditional $k_{\rm max}^{-1}\sim 10 h^{-1}$\ 
Mpc used in galaxy surveys.  In this case one might want the equations describing a general intrinsic alignment model; these are given in Appendix~\ref{app:tensor}.

\subsection{Galaxy power spectrum}
\label{ss:gps}

Having defined the intrinsic alignment power spectra, we now compute the observed redshift-space power spectrum of the galaxies, i.e. of Eq.~(\ref{eq:tdgx2}).  We 
consider here only scalar intrinsic alignments with even $l$ only (i.e. we assume inversion symmetry).  We 
first plug in Eq.~(\ref{eq:elm}),
\begin{equation}
\tdg(\bk) = (b+f\mu^2)\tdm(\bk) + \sum_{l\ge 2,\rm\,even} \tilde\epsilon_{l0}(\bk) Y_{l0}(\mR\ehat_3).
\end{equation}
Taking the power spectrum gives
\begin{eqnarray}
\Pg(\bk) \!\!\!\!&=&\!\!\!\! (b+f\mu^2)^2\Pm(k) 
\nonumber \\ &&\!\!\!\!
+ 2(b+f\mu^2) \Re \!\!\sum_{l\ge 2,\rm\,even}\! (-\rmi)^l \sqrt{\frac{4\pi}{2l+1}}\,\Pme^l(k) Y_{l0}(\mR\ehat_3)
\nonumber \\ &&\!\!\!\!
+  \sum_{ll',\rm\,even} \rmi^{l-l'} \frac{4\pi P_\epsilon^{ll'0}(k)
Y_{l0}^\ast(\mR\ehat_3)Y_{l'0}(\mR\ehat_3)}{\sqrt{(2l+1)(2l'+1)}}.
\label{eq:p1}
\end{eqnarray}
Here the first term is the conventional redshift-space power spectrum, the second comes from correlations of the matter (and velocity) fields with intrinsic alignments, and the final term is the pure intrinsic alignment term.

Equation~(\ref{eq:p1}) can be simplified by replacing $Y_{l0}$ with a Legendre polynomial.\footnote{Following convention, we denote the Legendre 
polynomials by $P_l$ and associated Legendre polynomials by $P_l^m$.  By context these should not be confused with power spectra.}  We can evaluate its argument via
\begin{equation}
\mu = \ehat_3\cdot\hat\bk = (\mR\ehat_3)\cdot(\mR\hat\bk) = (\mR\ehat_3)\cdot\ehat_3,
\end{equation}
where the first equality is the definition of $\mu$, the second is the definition of a rotation matrix, and the third follows from the condition $\mR\hat\bk=\ehat_3$.  Thus
\begin{equation}
\sqrt{\frac{4\pi}{2l+1}}\,Y_{l0}(\mR\ehat_3) = P_l(\mu).
\label{eq:Lpoly}
\end{equation}
This substitution gives rise to
\begin{eqnarray}
\Pg(\bk) \!\!&=& (b+f\mu^2)^2\Pm(k) 
\nonumber \\ &&
+ 2(b+f\mu^2) \sum_{l\ge 2,\rm\,even} (-1)^{l/2} \Pme^l(k) P_l(\mu)
\nonumber \\ &&
+ \sum_{ll'\rm\,even} (-1)^{(l-l')/2} P_\epsilon^{ll'0}(k) P_l(\mu)P_{l'}(\mu).
\label{eq:p2}
\end{eqnarray}

\section{Tidal alignment models}
\label{sec:t}

The best-motivated model for intrinsic alignments is to suppose that large-scale tidal fields would induce a preferential direction in galaxy 
formation.  The selection probability for a galaxy would then depend on the both $\nhat$ and the configuration of the tidal field.  Tidal fields from 
linear regime density fluctuations are small: by definition to be linear the tidal field is $\le H^{2}$ where $H$ is the Hubble constant, and hence 
must also be less than $t_{\rm d}^{-2}$ where $t_{\rm d}$ is the relevant dynamical timescale of the galaxy at any stage of its collapse.  Moreover, 
the small-scale nonlinear tidal fields are stronger than the large-scale tidal fields because $k^3\Pm(k)$ is a rapidly increasing function of $k$.  
A plausible model for tidal alignments would then be to take the large-scale tidal field as a perturbation and Taylor expand to lowest nonvanishing order:
\begin{equation}
\epsilon(\nhat|\bx) = \frac{A}{4\pi Ga^2\bar\rho_{\rm m}(a)} \left(\hat n_i \hat n_j \nabla_i \nabla_j - \frac13\nabla^2\right) \Psi(\bx),
\label{eq:tidal}
\end{equation}
where $\Psi(\bx)$ is the Newtonian gravitational potential, $A$ is the expansion coefficient, and the denominator $4\pi Ga^2\bar\rho_{\rm m}$ serves to 
make $A$ dimensionless.\footnote{Whether in practice a Taylor expansions such as this is appropriate for all types of galaxies is an open question.
For example, higher-order terms in the tidal field could dominate, or the actual tidal field could lie outside the radius of convergence of the series.}
The second derivative of $\Psi$ is the tidal field tensor, and the specified contraction with $\nhat$ is the only possibility 
consistent with the vanishing angle-average of $\epsilon(\nhat|\bx)$ since the tidal field is a quadrupole.  Note that even for the same population of galaxies, 
it is possible for $A$ to be 
redshift-dependent.  This dependence cannot be predicted without a detailed microphysical model of galaxy alignments.  Using the Poisson equation,
Eq.~(\ref{eq:tidal}) can 
also be re-expressed as
\begin{equation}
\epsilon(\nhat|\bx) = A s_{ij}(\bx) \hat n_i \hat n_j,
\label{eq:epshat}
\end{equation}
where
\begin{equation}
s_{ij} = \left( \nabla_i \nabla_j \nabla^{-2} - \frac13 \delta_{ij} \right)\dm(\bx)
\label{eq:sij}
\end{equation}
is the dimensionless tidal field.

An intrinsic alignment model of the form of Eq.~(\ref{eq:tidal}) could also be motivated by renormalization group arguments 
\citep{2009arXiv0902.0991M}: at first order in perturbation theory, the density $\delta$ and tidal field tensor $s_{ij}$ fully describe the history of 
any patch of material in the Universe.  Presumably second and third-order terms could be incorporated into Eq.~(\ref{eq:tidal}) and used to define a 
renormalized intrinsic alignment model in analogy to the \citet{2009arXiv0902.0991M} approach to the galaxy density. We note that for disk galaxies, an 
inclination-dependent selection effect combined with the quadratic tidal alignment model \citep{2001ApJ...559..552C, 2001MNRAS.320L...7C, 
2002MNRAS.332..788M} is an example of such a second-order term.  Like the second-order biasing of galaxies, quadratic alignment is 
most prominent on small (quasilinear) scales.  We briefly return to this issue in Section~\ref{ss:disk}. A full analysis of the quasilinear regime, 
including quadratic alignments and their effects on e.g. the galaxy bispectrum and quasilinear RSDs, is beyond the scope of this paper.

We now estimate the effect of tidal alignments on the observed galaxy power spectrum.  In Fourier space, Eq.~(\ref{eq:epshat}) becomes
\begin{equation}
\tilde\epsilon(\nhat|\bk) = A \left[(\nhat\cdot\hat\bk)^2 - \frac13\right]\tdm(\bk) = \frac23A P_2(\nhat\cdot\hat\bk)\tdm(\bk).
\end{equation}
Expanding in terms of spherical harmonics, this implies
\begin{equation}
\tilde\epsilon_{20}(\bk) = -\frac23A \tdm(\bk),
\label{eq:tidal20}
\end{equation}
with all other components equal to zero.  The implied power spectra are
\begin{equation}
\Pme^l(k) = -\frac23A\Pm(k)\delta_{l2}
\label{eq:tidal-me}
\end{equation}
and
\begin{equation}
P_\epsilon^{ll'm}(k) = \frac49A^2\Pm(k)\delta_{l2}\delta_{l'2}\delta_{m0}.
\label{eq:tidal-ee}
\end{equation}
The fact that only $l=2$ components appear is a direct result of the alignment model being linear in the tidal field, which transforms as a quadrupole.
Substitution into Eq.~(\ref{eq:p2}) gives
\begin{eqnarray}
\Pg(\bk) &=& (b+f\mu^2)^2\Pm(k) + \frac43A(b+f\mu^2)P_2(\mu)\Pm(k)
\nonumber \\ && + \frac49A^2[P_2(\mu)]^2\Pm(k).
\end{eqnarray}
Because $\epsilon$ is completely determined by $\dm$ in this model, this simplifies to
\begin{eqnarray}
\Pg(\bk) &=& \left[ b + f\mu^2 + \frac23AP_2(\mu) \right]^2\Pm(k)
\nonumber \\
&=& \left[ b-\frac A3 + (f+A)\mu^2 \right]^2\Pm(k).
\label{eq:tidal-pg}
\end{eqnarray}
We conclude that in the tidal alignment model, the observed redshift-space power spectrum retains the same functional form, but the intrinsic 
alignments alter the coefficients.  The ``obvious'' systematics tests for redshift space distortion measurements, such as testing the scale dependence 
of $\beta$ or the ``constant+$\mu^2$'' dependence of $\sqrt{\Pg(\bk)}$, will not detect this effect.  The intrinsic alignments change the normalization 
of the real-space ($\mu=0$) galaxy power spectrum, but this is degenerate with bias.  They also change the apparent rate of growth of structure, 
$f[\Pm(k)]^{1/2}\rightarrow (f+A)[\Pm(k)]^{1/2}$, which is a real systematic error whose fractional magnitude is $|A|/f$.  One can also note the sign 
of the effect: if galaxies are preferentially selected in regions where the tidal field is compressional along the line of sight ($A>0$) then the RSD 
is enhanced; whereas if galaxies are preferentially selected in regions where the tidal field is stretching along the line of sight ($A<0$) then the 
RSD is suppressed.

\section{Models for $A$}
\label{sec:micro}

Now that we have established the effect of tidal alignments on the galaxy power spectrum, it is time to consider the likely value of the $A$ parameter.  Since this parameter describes how tidal fields feed into galaxy orientations and ultimately observational selection criteria, we consider two separate cases.  The first is an ``elliptical galaxy'' case in which selection criteria are dependent on e.g. isophotal magnitudes or effective radius, which appear different depending on whether one looks down the long axis or the short axis of the elliptical.  The second is a ``disk galaxy'' case in which the selection criteria are blue continuum or emission line intensities that suffer inclination-dependent extinction.  In both models, we can roughly predict the normalization of how much galaxy orientations affect selection but not how much tidal fields affect orientation; for the latter, we take observational results on the correlations of galaxy ellipticities with large scale structure.

Note that all of the models considered here are merely examples and may not represent the actual contamination in a particular survey.  The results depend sufficiently strongly on the detailed selection criteria that each survey must do its own calculation, either along the lines of those presented here or using simulations.

\subsection{Elliptical galaxy}
\label{ss:ellip}

We model an elliptical galaxy as an optically thin (i.e. negligible dust) triaxial system.  The volume emissivity at a position $\bs$ relative to the centre of the 
galaxy is taken to be $j({\bs})={\cal J}(\rho)$, where the ellipsoidal radius $\rho$ is $\rho^2 = \bs\cdot{\mathbfss V}^{-1}\bs$.  The matrix ${\mathbfss V}$ is 
symmetric and unimodular ($\det{\mathbfss V}=1$) and contains information about the anisotropy of the galaxy, whereas ${\cal J}(\rho)$ specifies the radial profile.  
For a spherical galaxy, ${\mathbfss V}={\mathbfss I}$.  We define the deviation from spherical symmetry to be ${\mathbfss W}\equiv{\mathbfss V}-{\mathbfss I}$, and 
will work in the limit where ${\mathbfss W}$ is small.  To first order, ${\mathbfss W}$ is traceless.  It is described by 5 numbers: 2 linearly independent 
eigenvalues that specify the amount and type (oblate vs. prolate) of triaxiality, and 3 Euler angles.  We will argue here that for small ${\mathbfss W}$, the 
selection probability $\Upsilon({\mathbfss Q}\ehat_3,\bx)$ is proportional to $W_{33}(\bx)$.  Then in accordance with Eq.~(\ref{eq:epsilon}) we will find that 
$\epsilon(\ehat_3|\bx)\propto \langle W_{33}\rangle(\bx)$, where the coefficient of proportionality depends on the selection algorithm.  We will argue that $\langle 
W_{ij}\rangle(\bx) \propto s_{ij}(\bx)$ (see Appendix~\ref{app:Euler} for an explicit demonstration in terms of the integral over ${\mathbfss Q}$).  The coefficient 
of proportionality, needed to determine $A$, will be fixed using measurements of galaxy ellipticity-density correlations.

We do not observe the full 3-dimensional structure of the galaxy, but rather its 2-dimensional projection
\begin{equation}
I(\bs_\perp) = \int j(\bs_\perp + s_3\ehat_3) \,\rmd s_3,
\label{eq:i}
\end{equation}
where $\bs_\perp$ is a position in the plane of the sky, i.e. the 12-plane.

In Appendix~\ref{app:proj}, we show that the observed two-dimensional projection is
\begin{equation}
I(\bs_\perp) = \left(1+\frac12W_{33}\right) I_0(c).
\label{eq:i2}
\end{equation}
Here
\begin{equation}
I_0(c) \equiv \int {\cal J}\left(\left| {\bmath c}+s_3\ehat_3\right|\right) \,\rmd s_3
\end{equation}
is the projected image of the spherical galaxy model,
\begin{equation}
{\bmath c} = \left({\mathbfss I} - \frac12{\mathbfss W}_{\perp}\right)\bs_\perp
\label{eq:c}
\end{equation}
is a rescaled skewer position in the 12-plane, and ${\mathbfss W}_\perp$ is the projection of ${\mathbfss W}$ into the 12-plane (i.e. the $2\times 2$ submatrix).  Note that ${\mathbfss W}_\perp$ need not be traceless; indeed Tr$\,{\mathbfss W}_\perp=-W_{33}$.

The observed properties of this galaxy are as follows.  The effective radius $r_{\rm e}$ is reduced from its spherical version in accordance with Eq.~(\ref{eq:c}):
\begin{equation}
r_{\rm e} = \left(1 - \frac14W_{33} \right)r_{{\rm e}0}.
\label{eq:re}
\end{equation}
The total flux $F$ of the galaxy is unchanged.  Finally, the ellipticities of the isophotes (surfaces of constant $c$) components are
\begin{equation}
e_1 = \frac12(W_{11}-W_{22}) \;\;\;\;{\rm and}\;\;\;\;
e_2 = W_{12}.
\label{eq:ellip}
\end{equation}

LRGs are usually selected by a combination of cuts in colour space, combined with a magnitude cut.  As long as the galaxy is optically thin, there is 
likely to be very little orientation-dependent change in colour.  However, the apparent magnitude of a galaxy may depend strongly on its orientation.  
Here we provide some examples.

{\em Model magnitudes}: If galaxies are selected based on model magnitudes, and the model fit accurately describes the radial profile, then the 
model-fit magnitude corresponds to the orientation-independent flux $F$.  This gives $\epsilon=0$.

{\em Petrosian magnitudes}: For galaxies described by a single radial profile ${\cal J}(\rho)$ (e.g. a deprojected de Vaucouleurs model), that are 
well-resolved (in the sense that the Petrosian radius is well beyond the smearing of the central cusp due to the point-spread function), and have small 
ellipticity (${\mathbfss W}$ small), the Petrosian magnitude should be invariant under the rescaling of $r_{\rm e}$ described by Eq.~(\ref{eq:re}).  
This is because the Petrosian radius will be a fixed multiple of $r_{\rm e}$ and will capture a fixed fraction of the galaxy's flux.  Within these 
approximations we would have $\epsilon=0$, but the robustness of this conclusion should be tested in simulations if high accuracy is required.

{\em Aperture magnitudes}:
If galaxies are selected based on aperture magnitudes, then we count more of a galaxy's light when it is viewed down a long axis than a short axis.  As an example, suppose that the aperture is cut off at $y$ effective radii the spherical galaxy model, and the shape of the galaxy's profile is such that the fraction of the luminosity within $yr_{\rm e}$ is ${\cal Q}(y)$.  A 1 per cent decrease in $r_{\rm e}$ results in a 1 per cent increase in $y$ and hence a $y{\cal Q}'(y)/{\cal Q}(y)$ per cent increase in ${\cal Q}(y)$.  Then using Eq.~(\ref{eq:re}), the aperture flux $F_{\rm ap}$ is modified according to
\begin{equation}
\frac{ F_{\rm ap} }{ F_{\rm ap0} } = 1 + \frac14\frac{y{\cal Q}'(y)}{{\cal Q}(y)}W_{33}.
\end{equation}
If the galaxies are selected according to a flux limit and the cumulative luminosity function of galaxies has logarithmic slope $\eta$, i.e. $\rmd\ln \bar n/\rmd\ln F_{\rm min} = -\eta$, then a 1 per cent increase in measured flux for galaxies near threshold corresponds to an $\eta$ per cent increase in the number density.  Thus
\begin{equation}
\epsilon = \frac\eta4\frac{y{\cal Q}'(y)}{{\cal Q}(y)}\langle W_{33}\rangle.
\end{equation}

{\em Isophotal magnitudes}:
Isophotal magnitudes are similar to aperture magnitudes in that we count more of a galaxy's light when it is viewed down a long axis than a short axis.  However, in this case the cutoff for flux determination depends on the surface brightness profile.   The surface brightness at $y$ effective radii is
\begin{equation}
I(y) = \frac{F}{2\pi r_{\rm e}^2} \frac{{\cal Q}'(y)}y.
\end{equation}
If we change $r_{\rm e}$ but hold the surface brightness constant (as appropriate for determining how the isophotes move), we find
\begin{equation}
0 = \frac{\delta I(y)}{I(y)} = -2\frac{\delta r_{\rm e}}{r_{\rm e}} + \left[\frac\rmd{\rmd y}\ln\frac{{\cal Q}'(y)}y\right]\delta y.
\end{equation}
The quantity in brackets evaluates to ${\cal Q}''(y)/{\cal Q}'(y)-y^{-1}$.  We also know from Eq.~(\ref{eq:re}) that $\delta r_{\rm e}/r_{\rm e}=-W_{33}/4$.  Thus we may solve for $\delta y$:
\begin{equation}
\delta y = \frac{-W_{33}}{2[{\cal Q}''(y)/{\cal Q}'(y)-y^{-1}]}.
\end{equation}
This implies a fractional change in isophotal flux of
\begin{equation}
\frac{ F_{\rm is} }{ F_{\rm is0} } = 1 + \frac{{\cal Q}'(y)\delta y}{{\cal Q}(y)}
= 1 - \frac{{\cal Q}'(y)W_{33}}{2{\cal Q}(y)[{\cal Q}''(y)/{\cal Q}'(y)-y^{-1}]}.
\end{equation}
For a luminosity function slope $\eta$, we find
\begin{equation}
\epsilon = - \frac{\eta{\cal Q}'(y)}{2{\cal Q}(y)[{\cal Q}''(y)/{\cal Q}'(y)-y^{-1}]} \langle W_{33}\rangle.
\end{equation}
Note that the coefficient of $W_{33}$ is positive, because ${\cal Q}''(y)/{\cal Q}'(y)-y^{-1}<0$ for any surface brightness profile that decreases as one moves outward (e.g. de Vaucouleurs).

\begin{figure}
\includegraphics[angle=-90,width=3.2in]{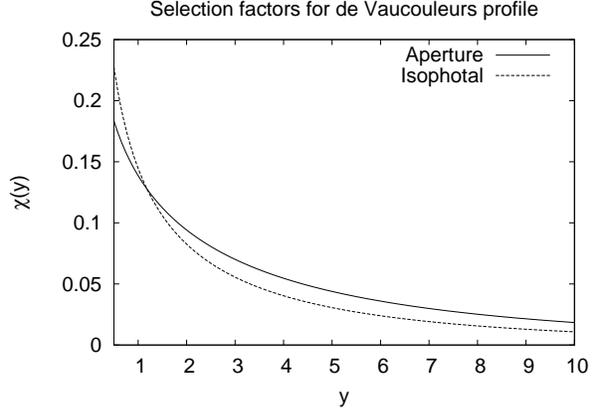}
\caption{\label{fig:dev}The selection factors $\chi$ of Eq.~(\ref{eq:ew}) for a de Vaucouleurs galaxy.  Either aperture or isophotal magnitudes are 
used for the flux cut.  The factor depends on $y$, the number of effective radii at which the photometry is cut off for a typical galaxy at the faint 
end of the sample.  If model or Petrosian magnitudes were used, we would have $\chi=0$.} \end{figure}

In all of these cases, one may write
\begin{equation}
\epsilon = \eta\chi \langle W_{33}\rangle,
\label{eq:ew}
\end{equation}
where $\chi$ depends on the method of computing fluxes (model, Petrosian, aperture, or isophotal), the galaxy profile, and (in the latter two cases) 
the typical number of effective radii $y$ at which the flux is computed for galaxies near the threshold.  For the \citet{1948AnAp...11..247D} profile, 
where ${\cal Q}'(y)\propto y\exp(-7.67y^{1/4})$, values of $\chi$ are shown in Figure~\ref{fig:dev}.  Note, however, that Petrosian, aperture, or 
isophotal magnitudes of galaxies at radii affected by the point-spread function of the telescope would exhibit more complicated behavior that depends also on the 
dimensionless resolution factor (model magnitudes will be affected too if the model does not correctly describe the galaxy).

We now suppose that the mean value of ${\mathbfss W}$ for galaxies in a particular region of space has some dependence on the tidal field surrounding it.
Since both $W_{ij}$ and the tidal field $s_{ij}$ are traceless-symmetric tensors and have quadrupolar symmetry, the lowest-order allowed term in the Taylor expansion of $W_{ij}$ as a function
of $s_{ij}$ is:
\begin{equation}
\langle W_{ij}\rangle = 2Bs_{ij} = 2B \left(\nabla_i\nabla_j\nabla^{-2} - \frac13\right)\dm(\bx).
\label{eq:Wij}
\end{equation}
An alternative argument for Eq.~(\ref{eq:Wij}) in terms of the Euler angles can be found in Appendix~\ref{app:Euler}.  We also argue there that although the full specification 
of the alignment of a triaxial galaxy to linear order in $s_{ij}$ requires 2 parameters, the statistical effect on $\langle W_{ij}\rangle$ only involves 1 
parameter ($B$).

Comparing Eq.~(\ref{eq:Wij}) to Eqs.~(\ref{eq:epshat}) and (\ref{eq:ew}) gives the proportionality coefficient,
\begin{equation}
A = 2\eta\chi B.
\label{eq:AB}
\end{equation}
As discussed in Sec.~\ref{ss:ia}, there is an effect on large scale structure only when the galaxies have nonrandom orientations ($B\neq 0$) and when there is an 
orientation-dependent selection effect ($\chi\neq 0$).

The normalization constant $B$ depends on the details of elliptical galaxy formation.  Rather than choosing a value based on theoretical considerations, we normalize $B$ to measurements of intrinsic ellipticity correlations of elliptical galaxies.  Using Eq.~(\ref{eq:ellip}), we find that for a Fourier mode $\bk$ perpendicular to the line of sight, e.g. in the 1-direction ($\bk=k\ehat_1$), we have $\tilde e_1(\bk) = B\tdm(\bk)$.  This implies a cross-power spectrum between matter and ellipticity of
\begin{equation}
P_{\rm me}(k) = B\Pm(k).
\end{equation}
This power spectrum -- or more precisely, the cross-correlation -- was measured by \citet{2007MNRAS.381.1197H}, who were interested in intrinsic alignments as a contaminant of cosmic shear measurements.  For the latter reason, \citet{2007MNRAS.381.1197H} presented not the correlation with ellipticity components $e_{1,2}$ but with ``intrinsic shear'' $\gamma^{\rm I}_{1,2}\equiv e_{1,2}/1.74$, where 1.74 is the shear responsivity.  Note that in the notation of \citet{2008arXiv0808.3400B}, $b_\kappa = B/1.74$.  The projected correlation function $w_{\delta+}(r_{\rm p})$ of the matter and intrinsic shear is
\begin{equation}
w_{\delta+}(r_{\rm p}) = -\frac{b_\kappa}{2\pi} \int \Pm(k) J_2(kr_{\rm p}) k\,\rmd k.
\label{eq:wd}
\end{equation}
\citet{2007MNRAS.381.1197H} showed that the scale dependence implied by Eq.~(\ref{eq:wd}) is a good fit to the intrinsic alignments of LRGs.  They normalized their correlation function by its value at $r_{\rm p}=20h^{-1}\,$Mpc and $z=0.3$, and using the nonlinear matter power spectrum \citep{2003MNRAS.341.1311S}, Eq.~(\ref{eq:wd}) gives a correlation function of $-3.3b_\kappa h^{-1}\,$Mpc.  Comparing to the observed value of $0.059(L/L_0)^{1.48}h^{-1}\,$Mpc (where $L$ is the luminosity $K+e$-corrected to $z=0$ and $L_0$ corresponds to a corrected $r$-band absolute magnitude $M_r^{0.0}$ of $-22$) suggests that
\begin{equation}
b_\kappa = (-0.018\pm0.006)\left(\frac L{L_0}\right)^{1.48\pm 0.64}.
\end{equation}
Thus at $z = 0.3$, $B\approx -0.03$ for LRGs of magnitude $M_r^{0.0}= -22$, rising to $-0.06$ at $M_r^{0.0}=-22.5$ and $-0.12$ at $M_r^{0.0}=-23$.

In order to estimate the anisotropic selection parameter $A$, we additionally need the slope of the cumulative luminosity function $\eta = -\rmd\ln\bar 
n/\rmd\ln F_{\rm min}$.  We may use the LRG luminosity function of \citet{2006MNRAS.372..537W}.\footnote{\citet{2006MNRAS.372..537W} note that for 
typical LRGs, $M_r^{0.2} \approx M_r^{0.0} + 0.11$; but note that the absolute magnitude in \citet{2007MNRAS.381.1197H} is referenced to 10$h^{-1}$ pc, 
whereas \citet{2006MNRAS.372..537W} refers absolute magnitudes to 10 pc assuming $h=0.7$.  This introduces an additional offset of $5\log_{10}0.7$ so 
that the magnitudes in \citet{2006MNRAS.372..537W} are given by $M_r^{\rm Wake}=M_r^{0.0}-0.66$.} At $M_r^{0.0}=-22.5$, we find $\eta=4.0$.  If 
galaxies are selected with isophotal magnitudes with a surface brightness cutoff at $\sim 3r_{\rm e}$, then $\chi\sim 0.05$.  Combining with $B\sim 
-0.06$, this leads to $A\sim -0.024$.  Since at $z=0.3$, $f\approx 0.65$, we find that the fractional contamination to the redshift-space distortion 
measurement $|A|/f$ is $\sim 4$ per cent.  For LRGs a half a magnitude brighter at $M_r^{0.0}=-23$, $\eta=4.5$ and $B\sim-0.12$, so the contamination 
increases to $\sim 8$ per cent.  If one used aperture magnitudes instead, with a typical aperture cutoff at $\sim 3r_{\rm e}$, then we would have 
$\chi\sim 0.06$ and our estimated fractional contamination $|A|/f$ for these two cases would rise to 5 and 10 per cent, respectively.

These values are extremely rough and are merely intended to show that there exist regimes in which the cosmological interpretation of the 
redshift-space distortions is significantly altered.  The level of contamination for a particular survey depends in great detail on its selection 
criteria; if it is significant, $\eta\chi$ would best be determined by a combination of Monte Carlo simulations and direct use of the observed 
distribution of galaxies in colour-magnitude space rather than the simplified analytic arguments used here.  Also note the sign: for isophotal or 
aperture magnitude selection, where galaxies are more likely to pass cuts if aligned in the radial direction, we have $A<0$.

The leading spectroscopic LRG surveys today, such as the SDSS LRG survey \citep{2001AJ....122.2267E} and 2SLAQ \citep{2006MNRAS.372..425C}, use 
Petrosian or model magnitudes for their selection.  This dramatically reduces the intrinsic alignment contamination since the selection should be 
almost orientation-independent (except for discrepancies between the model and the actual galaxy profile, and for the handful of objects with 
significant dust).

\subsection{Disk galaxy}
\label{ss:disk}

We now consider an alternative model for anisotropic selection: a disk galaxy selected in either emission lines (e.g. [O{\sc\ ii}] or H$\alpha$) or 
optical/ultraviolet continuum will suffer less extinction if viewed face-on.  Therefore selection based on the apparent magnitude or emission line flux 
favors face-on galaxies.  The flux from the galaxy depends on the detailed distribution of stars or star-forming regions and dust.  For some 
applications the extinction and scattering problems can be reduced by working in the rest-frame infrared, however in large redshift surveys that 
require blue or ultraviolet continuum as part of the colour selection criteria (e.g. WiggleZ; \citealt{2007ASPC..379...72G}) or require a bright 
emission line to obtain a redshift (e.g. the Joint Dark Energy Mission, JDEM\footnote{URL: \tt http://jdem.gsfc.nasa.gov/}) this is not an option.  In 
this section, we will first relate the distribution of disk orientations to $b_\kappa$ and then derive the $A-b_\kappa$ relation for general 
inclination-dependent flux $\Phi(i)$.  We will then consider a few toy models of $\Phi(i)$ and their implications.  More complex models have been 
considered elsewhere (e.g. \citealt{1989MNRAS.239..939D, 1994AJ....107.2036G, 1995AJ....110.1059G}).

We consider a distribution of disks whose normal vectors $\Lhat$ have probability
\begin{equation}
P(\Lhat) = \frac1{4\pi} \left(1 + Bs_{ij}\hat L_i \hat L_j \right),
\label{eq:plhat}
\end{equation}
where $B$ describes the extent to which large-scale tidal fields affect the orientation of galactic disks.  The anisotropy of the form 
Eq.~(\ref{eq:plhat}) is the only correction that is allowed by symmetry to first order in the large-scale tidal field.  Such an anisotropic alignment can occur due to nonlinear evolution in tidal torque models \citep{2008ApJ...688..742H}.  For a geometrically thin axisymmetric disk perpendicular to $\Lhat$, the inclination is $\cos i = \hat L_3$ and the axis ratio is $b/a=\cos i$.  Thus the ellipticity is
\begin{equation}
e = \frac{1-(b/a)^2}{1+(b/a)^2} = \frac{1-\hat L_3^2}{1+\hat L_3^2} = \frac{\hat L_1^2 + \hat L_2^2}{1+\hat L_3^2}.
\end{equation}
The position angle $\phi$ of the apparent major axis is $\pi/2$ away from the position angle of the projected angular momentum vector, $\tan^{-1}(\hat L_2/\hat L_1)$.  Thus the components $e_1=e\cos2\phi$ and $e_2=e\sin2\phi$ of the ellipticity tensor are
\begin{equation}
e_1 = -\frac{\hat L_1^2 - \hat L_2^2}{1+\hat L_3^2} \;\;\;\;{\rm and}\;\;\;\;
e_2 = -\frac{2\hat L_1\hat L_2}{1+\hat L_3^2}.
\end{equation}
The mean ellipticity can be obtained by integrating the probability distribution Eq.~(\ref{eq:plhat}) for $\Lhat$ over the unit sphere.  That is,
\begin{eqnarray}
\langle e_1 + \rmi e_2 \rangle \!\! &=& \!\!-\frac1{\int_{S^2} p(i) \rmd^2\hat L}
\nonumber \\ && \!\!\times
\int_{S^2} \frac{(\hat L_1+\rmi\hat L_2)^2}{1+\hat L_3^2}
(1 + Bs_{ij}\hat L_i \hat L_j)p(i) \rmd^2\hat L
,\nonumber \\ &&
\end{eqnarray}
where $p(i)$ is the selection probability for a galaxy at inclination $i$.  We switch to spherical polar coordinates ($i,\phi$), and performing the $\phi$ integration we obtain
\begin{equation}
\langle e_1\rangle = -K (s_{11}-s_{22}); \;\;\;\;\langle e_2\rangle = -2Ks_{12},
\label{eq:e12}
\end{equation}
where
\begin{equation}
K = \frac{\int_0^\pi \sin^5i\, (1+\cos^2i)^{-1}p(i)\,\rmd i}{4\int_0^\pi p(i)\sin i\,\rmd i}.
\label{eq:K}
\end{equation}

If the anisotropic selection is a small effect so that $p(i)\approx\,$constant, then we have $K=\pi/4-2/3\approx 0.119$.\footnote{This is obtained by substituting $x=\cos i$ and decomposing in partial 
fractions.}  In models with large anisotropic selection effects, $K$ might be decreased -- for example, in the extreme example below of an optically thick uniform disk whose apparent luminosity is 
proportional to $\cos i$, if the slope of the luminosity function is $\eta=\rmd\ln\bar n/\rmd\ln F_{\rm min}=2$ then $p(i)\propto\cos^2i$.  This would result in $K=0.044$, nearly a factor of 3 smaller.  
We will parameterize this uncertainty with the parameter $K_{\rm a}=K/0.119$, which is equal to unity for isotropically selected thin disks and declines if face-on disks are preferentially 
selected.\footnote{By inserting $\delta$-functions in $p(i)$, we see that $K_{\rm a}\rightarrow 0$ if only face-on galaxies are selected, and $K_{\rm a}\rightarrow (\pi-8/3)^{-1}\approx2.11$ if only 
edge-on disks are selected.} Note also that the effective $K$ would be decreased if we took into account that even a perfectly edge-on disk does not have $e=1$ because of its finite thickness and the 
contribution of the bulge.  Both effects imply that in practice one should have $K_{\rm a}<1$.

For a Fourier mode $\bk\parallel\ehat_1$ in the plane of the sky, we find
$\tilde e_1(\bk) = KB\tdm(\bk)$, and hence
\begin{equation}
b_\kappa = -\frac{KB}{1.74} = -0.068K_{\rm a}B.
\end{equation}
This establishes the relation between the disk orientations and the $b_\kappa$ parameter of \citet{2008arXiv0808.3400B}.

In order to proceed, we must determine how the selection probability $\epsilon$ is related to $B$.  We assume a flux function with slope $\rmd\ln \bar n/\rmd\ln F_{\rm min}=-\eta$.  Then the number density of galaxies per logarithmic range in intrinsic flux $F_{\rm i}$ per unit solid angle of orientation is
\begin{equation}
{\cal N}(F_{\rm i},\Lhat) \propto F_{\rm i}^{-\eta}\left(1 + Bs_{ij}\hat L_i \hat L_j \right).
\end{equation} 
The observed flux is $F=F_{\rm i}\Phi(i)$.  The number density of galaxies above some threshold flux $F_0$ is then
\begin{equation}
N(>F_0) \propto \int \rmd^2\Lhat \int_{F_0/\Phi(i)}^\infty \rmd\ln F_{\rm i}\,F_{\rm i}^{-\eta}\left(1 + Bs_{ij}\hat L_i \hat L_j \right).
\end{equation}
One can separate the integral over $\rmd^2\Lhat$ into an integral over inclination $\sin i=\hat L_3$ and position angle $\phi$.  The position angle integrates out trivially,
\begin{eqnarray}
N(>F_0) \!\!\!\! &\propto&\!\!\!\! \int_0^\pi \sin i\,\rmd i\,
\int_{F_0/\Phi(i)}^\infty \rmd\ln F_{\rm i}\,F_{\rm i}^{-\eta}
\nonumber \\ && \!\!\!\!\times
\left[1 + Bs_{33}\cos^2i + \frac B2(s_{11}+s_{22})\sin^2i \right].
\end{eqnarray}
Using the tracelessness of $s_{ij}$ to simplify the last term, and performing the $F_{\rm i}$ integral, we get
\begin{equation}
N(>F_0) \propto \int_0^\pi [\Phi(i)]^\eta
\left[1 + Bs_{33}P_2(\cos i)\right] \sin i\,\rmd i.
\end{equation}
The anisotropic dependence is that due to $s_{33}$:
\begin{equation}
\epsilon = Bs_{33} \frac{\int_0^\pi [\Phi(i)]^{\eta} P_2(\cos i)\sin i\,\rmd i}
{\int_0^\pi [\Phi(i)]^{\eta} \sin i\,\rmd i}.
\end{equation}
Thus, defining
\begin{equation}
\psi \equiv -\frac{\int_0^\pi [\Phi(i)]^{\eta} P_2(\cos i)\sin i\,\rmd i}
{\int_0^\pi [\Phi(i)]^{\eta} \sin i\,\rmd i},
\label{eq:psi-int}
\end{equation}
we see from Eq.~(\ref{eq:epshat}) that $A=-\psi B$ or
\begin{equation}
A = 14.6 \frac\psi{K_{\rm a}} b_\kappa.
\label{eq:ABd}
\end{equation}
Since $P_2(\cos i)$ is bounded between $-\frac12$ and $1$, it follows that $-\frac12\le\psi\le1$; for selection in favour of face-on galaxies $\psi>0$.  
The parameter $K_{\rm a}$ is obtained from Eq.~(\ref{eq:K}) using the selection probability, which is $p(i)\propto[\Phi(i)]^\eta$.

In order to use Eq.~(\ref{eq:ABd}) we need to construct a model for the angular distribution of emitted radiation $\Phi(i)$.  We consider several toy examples.  These are simple plane-parallel models 
in which we consider the dust optical depth $\tau$ through the galactic disk.  This need not equal the optical depth relevant for extinction-correcting the stellar continuum or emission lines.  For 
example, in an [O{\sc\ ii}] or H$\alpha$ flux-limited survey where there may be significant internal extinction in H{\sc\ ii} regions, the analysis below should include only the diffuse contribution 
due to propagation from the disk (which is affected by disk inclination), rather than the ``total'' optical depth.

{\em Uniform slab}: We suppose the emitters have the same vertical distribution as the diffuse dust, and the vertical optical depth through the disk is $\tau$.
In this case the observed optical depth 
at inclination $i$ is $\tau|\sec i|$.  Then $\Phi(i)$ is proportional to the probability of a photon emitted at random location in the disk reaching us without being absorbed, i.e. $\Phi(i) = 
(1-\rme^{-\tau|\sec i|})/(\tau|\sec i|)$.  The integrals in Eq.~(\ref{eq:psi-int}) are not represented by any commonly used function and must be evaluated numerically, except in the limiting case of an
optically thick slab ($\tau\gg1$).

{\em Uniform, optically thick slab}: If the disk is optically thick to dust absorption and has emitters (young stars or H{\sc\ ii} regions) with the 
same vertical distribution as the diffuse dust, then the surface brightness of the disk is inclination-independent.  This gives a flux proportional to 
projected area or $\Phi(i)\propto|\cos i|$.  Both the numerator and denominator in Eq.~(\ref{eq:psi-int}) are then polynomials in $\cos i$; the integral evaluates in closed form to
\begin{equation}
\psi = \frac\eta{\eta+3}.
\end{equation}
There is no closed-form analytic expression for $K_{\rm a}$ for general $\eta$.  However for integer $\eta$ one may use the substitution $x=\cos i$ and 
integrate via partial fractions to get
\begin{equation}
K = \left\{\begin{array}{lll}
\frac\pi4-\frac23 & & \eta=0 \\
\ln2-\frac58 & & \eta=1 \\
\frac{12}5-\frac34\pi & & \eta=2 \\
\frac{17}{12}-2\ln2 & & \eta=3 \\
\frac54\pi - \frac{82}{21} & & \eta=4
\end{array}\right.,
\end{equation}
which imply $K_{\rm a}=1$, 0.574, 0.369, 0.256, and 0.187, respectively.  

{\em Emitting sheet embedded in absorbing slab}: If the disk has a thin emitting layer sandwiched between two absorbing slabs each of optical depth $\tau/2$, then $\Phi(i)\propto \rme^{-\tau|\sec 
i|/2}$.  With the substitution of $x=\cos i$, Eq.~(\ref{eq:psi-int}) is transformed into an 
exponential integral (Eq.~5.1.4 of \citealt{1972hmfw.book.....A}), giving
\begin{equation}
\psi = \frac{3E_4(\eta\tau/2)}{2E_2(\eta\tau/2)}-\frac12
\rightarrow \left\{ \begin{array}{ll}
-\frac14\eta\tau\ln(2.42\eta\tau) & \eta\tau\ll 1
\\
1 & \eta\tau\gg 1
\end{array}\right..
\end{equation}
The asymptotic expansions use Eqs.~(5.1.12) and (5.1.52) of \citet{1972hmfw.book.....A}.

{\em Power-law model}: \citet{1994AJ....107.2036G} estimated that the total (exponential model) magnitude of local Sc galaxies suffers an inclination-dependent correction $\Delta 
M=-\gamma\log_{10}(a/b)$ in the $I$ band, where $\gamma=1.02\pm0.08$ was found by minimizing the $\chi^2$ of the corrected Tully-Fisher relation fit.  
This implies $\Phi(i)\propto|\cos^{0.4\gamma}i|$, and
\begin{equation}
\psi = \frac{0.4\eta\gamma}{0.4\eta\gamma+3}.
\end{equation}


Note that in all of these cases additional corrections apply if isophotal or aperture magnitudes are used in place of exponential model fits \citep{1994AJ....107.2036G}.

\begin{figure}
\includegraphics[angle=-90,width=3.3in]{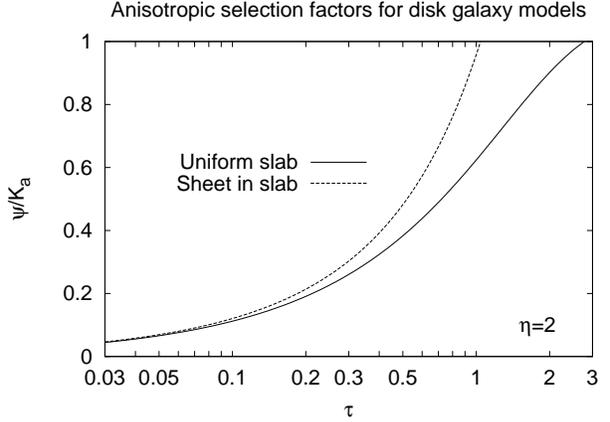}
\caption{\label{fig:disk}The orientation-dependent selection factor $\psi/K_{\rm a}$ for both the uniform slab and sheet-in-slab models for $\eta=2$, as a function of disk optical depth.
}
\end{figure}

We consider a slope of the cumulative luminosity function $\eta=-\rmd\ln\bar n/\rmd\ln F_{\rm min}\approx 2$.  This is roughly appropriate for a 
JDEM-type survey using the H$\alpha$ emission line to a density of several$\times 10^{-4}\,$Mpc$^{-3}$ at $z\sim 1$
\citep{1999ApJ...519L..47Y, 2009arXiv0902.2064S}
and $z\sim 2$ \citep{2008ApJS..175...48R, 2008MNRAS.388.1473G}.
In this case, the optically thick slab model would predict $\psi=0.40$ and $K_{\rm 
a}=0.37$, or $\psi/K_{\rm a}=1.1$.  The 
sheet-in-slab model predicts $\psi = 0.10$ and $K_{\rm a}=0.8$ or $\psi/K_{\rm a}=0.12$ at $\tau=0.1$, rising to $\psi/K_{\rm a}=0.5$ at $\tau=0.5$.  
More values are plotted in Figure~\ref{fig:disk}.  For the power-law model with $\gamma\approx1.0$ appropriate for local Sc galaxies in the $I$ band 
\citep{1994AJ....107.2036G}, we find $\psi=0.21$ and $K_{\rm a}=0.63$, or $\psi/K_{\rm a}=0.33$; a somewhat larger value might be expected at bluer 
wavelengths since there should be more internal extinction in the galaxy, and one might expect younger stars and H{\sc\,ii} regions to have a smaller 
scale height.  It is readily apparent that many models with significant optical depth give $\psi/K_{\rm a}$ in the range of a few tenths.

The last ingredient needed to make a prediction for $A$ is a prediction for or measurement of $b_\kappa$.  Unlike the case of elliptical galaxies, for 
disk galaxies there is at present no clear consensus on whether $b_\kappa\neq0$. Theoretically, $b_\kappa$ is exactly zero to lowest order in tidal 
torque theory because a collapsing protogalaxy requires not just a tidal field to spin up but also an anisotropic quadrupole moment of the mass 
distribution; the combination leads to alignments that do not correlate with the density perturbations and go rapidly to zero in the linear regime 
\citep{2001ApJ...559..552C, 2001MNRAS.320L...7C, 2002MNRAS.332..788M, 2004PhRvD..70f3526H}.  However, \citet{2008ApJ...688..742H} showed that when 
nonlinear evolution is considered, there is a squeezed bispectrum configuration of small-scale tidal fields, quadrupole moments of collapsing 
structures, and linear-scale density perturbations, so that tidal torque theory predicts $b_\kappa>0$.  \citet{2006MNRAS.371..750H} used $N$-body 
simulations to search for correlations of spin axes with the density (technically the cosmic shear) field and found none.

On the observational side, some early works found evidence for disk galaxy alignments with large-scale structure \citep{2002ApJ...567L.111L, 
2004ApJ...613L..41N} and interpreted their results as supportive of tidal torque theory.  On the other hand, studies with the much larger SDSS samples 
of blue galaxies have turned up no evidence for density-ellipticity correlations \citep{2007MNRAS.381.1197H, 2008arXiv0811.1995F}, even while 
confirming previous detections of LRG alignments \citep{1982A&A...107..338B} at high significance.  There have also been several claimed detections of 
disk galaxy orientations with the intermediate axis of the tidal field or with void shells, as would occur in tidal torque theory 
\citep{2002ApJ...567L.111L, 2006ApJ...640L.111T, 2007ApJ...671.1248L}.  These were all based on 3-dimensional modeling of the tidal field in which the 
line of sight is a preferred axis for both the tidal field (due to redshift space effects) and the galaxy angular momentum reconstruction; for this 
reason it is desirable to confirm these results with statistics such as position angle correlations where no direction is preferred by the algorithm.  
\citet{2007ApJ...670L...1L} searched for position angle correlations of neighbouring blue galaxies in SDSS; they observed a $3\sigma$ positive 
correlation in their innermost radial bin (3D redshift-space separation $s<3h^{-1}\,$Mpc) with nondetection at $s>3h^{-1}\,$Mpc. 
\citet{2007ApJ...670L...1L} noted that this scale dependence is consistent with a quadratic tidal alignment model, although the statistical 
significance is marginal.  This would lead to a next-order term involving $s_{ij}s_{kl}$ in Eq.~(\ref{eq:epshat}), whose correlations drop off rapidly 
at large scales.  This additional term would not affect RSDs in the large-scale limit, but the quasilinear regime would be affected.

Faced with these uncertainties, we consider as a worst-case scenario for the linear $b_\kappa$ coefficient for disk galaxy alignments the 
``pessimistic'' models from 
\citet{2007MNRAS.381.1197H}.  The pessimistic model corresponds to the $2\sigma$ upper limit on alignments of $\sim L_\star$ late-type galaxies; this 
is $|w_{\delta +}(r_{\rm p}=20h^{-1}\,{\rm Mpc},z=0.3)|=0.028h^{-1}\,$Mpc, or $|b_\kappa| = 0.008$.  Assuming $\psi/K_{\rm a}\sim 0.33$ in accordance 
with the power-law ($\gamma=1$) model, this leads to $|A|\sim 0.039$ and a fractional contamination to the redshift-space distortion $|A|/f$ of 6 per 
cent. This should not be taken to mean that there {\em is} contamination at this level, but rather as an example that may be appropriate if disk galaxy 
alignments are near present upper limits.  The true value of $|A|/f$ may be much smaller (although larger values at high redshift, or larger 
$\psi/K_{\rm a}$ for some types of selection, are possible).

\section{Redshift-space distortions versus cosmic shear}
\label{sec:cosmic}

It is of interest to compare the effect of tidal galaxy alignments on RSDs to their better-known effect on CS measurements, assuming that one uses similar samples of galaxies.  We will conclude that 
for the disk galaxies, even if $\psi/K_{\rm a}$ is of order unity, then for low-redshift measurements the cosmic shear is more affected than the RSDs.  However at redshifts of order unity, the 
fractional contamination to the RSDs is generally $\psi/K_{\rm a}$ times the fractional contamination to the cosmic shear.  Since we have seen that for many simple disk models $\psi/K_{\rm a}$ is a few 
tenths, and since future galaxy surveys designed to probe dark energy will be exploring the $z\sim 1$ range, tidal alignments should if uncorrected be considered a similar source of error for both 
types of measurements.

The contamination of the RSD measurement is very simply
\begin{equation}
\Delta\ln[f(z)G(z)] = \frac{A(z)}{f(z)} \approx \frac{A(z)}{[\Omega_{\rm m}(z)]^{0.6}}.
\label{eq:contam-z}
\end{equation}

The contamination of the CS signal is more complicated.  We consider the shear power spectrum for a broad redshift distribution, which for the pure cosmic shear case is given by
\begin{equation}
C_\ell^{\rm GG} = \int \frac{[W(r)]^2}{r^2}\Pm\left(k=\frac\ell r;\tau_0-r\right)\,\rmd r,
\end{equation}
where $\Pm$ is the matter power spectrum at the wavenumber $k$ and conformal time $\tau_0-r$, the conformal time today is $\tau_0$, the lensing window function is
\begin{equation}
W(r) = \frac32\Omega_{\rm m}H_0^2[1+z(r)] r \int_r^\infty \frac{r_{\rm s}-r}{r_{\rm s}} n(r_{\rm s})\,\rmd r_{\rm s},
\end{equation}
$n(r_{\rm s})$ is the probability distribution for sources to lie at distance $r_{\rm s}$ and we have assumed a flat universe (e.g. \citealt{2004PhRvD..70f3526H}).  Now suppose the galaxies have an intrinsic alignment described by the \citet{2008arXiv0808.3400B} parameter $b_\kappa$ and (as appropriate for large scales where the intrinsic alignments are due to the local tidal field) $P_{\delta,\tilde\gamma^{\rm I}}(k)=b_\kappa \Pm(k)$.  Then
the leading-order contamination in $b_\kappa$ is the ``GI'' or interference term of \citet{2004PhRvD..70f3526H}:
\begin{equation}
C_\ell^{\rm GI} = 2 \int \frac{n(r)W(r)}{r^2} b_\kappa \Pm\left(k=\frac\ell r;\tau_0-r\right)\,\rmd r.
\end{equation}

The evaluation of the fractional contamination to the lensing power spectrum $C_\ell^{\rm GI}/C_\ell^{\rm GG}$ requires a numerical computation in general.  However for conceptual purposes it helps to take the limit of nearby sources, i.e. $z\ll 1$ so that $r\approx H_0^{-1}z$, and take $\Pm(k)$ to be a power law $\Pm(k) = {\cal N}k^{\nseff}$.  For definiteness we take a source distribution
\begin{equation}
n(r_{\rm s}) = \frac{r_{\rm s}^2}{2r_\star^3}\rme^{-r_{\rm s}/r_\star},
\end{equation}
which implies a lensing window function
\begin{equation}
W(r) = \frac34\Omega_{\rm m}H_0^2r
\left(\frac r{r_\star}+2\right)\rme^{-r/r_\star}.
\end{equation}
The pure lensing contribution to the power spectrum is
\begin{equation}
C_\ell^{\rm GG} = \frac9{16}\Omega_{\rm m}^2H_0^4 {\cal N} r_\star^{1-\nseff} \ell^{\nseff} \int u^2(u+2)^2\rme^{-2u} u^{-\nseff} \rmd u
\end{equation}
with the substitution $u=r/r_\star$; this simplifies to
\begin{eqnarray}
C_\ell^{\rm GG} &=& \frac9{16}\Omega_{\rm m}^2H_0^4 {\cal N} r_\star^{1-\nseff} \ell^{\nseff}
\left( \frac14n_{\rm s}^{\rm eff\,2} - \frac{15}4\nseff + 13 \right)
\nonumber \\ && \times
\frac{\Gamma(3-\nseff)}{2^{3-\nseff}}.
\end{eqnarray}
Finally the GI contribution is
\begin{equation}
C_\ell^{\rm GI} = \frac34 \Omega_{\rm m}H_0^2 b_\kappa {\cal N} r_\star^{-1-\nseff} \ell^{\nseff}
\int u^3(u+2) \rme^{-2u} u^{-\nseff}\,\rmd u,
\end{equation}
which simplifies to
\begin{eqnarray}
C_\ell^{\rm GI} &=& \frac34\Omega_{\rm m}H_0^2 b_\kappa {\cal N} r_\star^{-1-\nseff} \ell^{\nseff}
\left( 4 - \frac12\nseff \right)
\nonumber \\ && \times
\frac{\Gamma(4-\nseff)}{2^{4-\nseff}}.
\end{eqnarray}
The ratio is
\begin{equation}
\frac{C_\ell^{\rm GI}}{C_\ell^{\rm GG}} =
\frac{4(3-\nseff)(8-\nseff)}{3(52-15\nseff+n_{\rm s}^{\rm eff\,2})}
\frac {b_\kappa}{\Omega_{\rm m}H_0^2r_\star^2}.
\end{equation}
The prefactor is a combination of dimensionless integrals that depends on $\nseff$; for typical values of the spectral index in the quasilinear regime 
$\nseff\approx-1.3$, it is $0.73$.  Also the median redshift is $z_{\rm med}\approx 2.67H_0^{-1}r_\star$.  This implies
\begin{equation}
\frac{C_\ell^{\rm GI}}{C_\ell^{\rm GG}} \approx 5.2\frac{b_\kappa}{\Omega_{\rm m}z_{\rm med}^2}.
\end{equation}
The fractional error in $\sigma_8$ depends on how $C_\ell$ scales with $\sigma_8$.  In the linear regime, $C_\ell^{\rm GG}\propto\sigma_8^2$, but in practical cases, CS uses the quasilinear regime and the actual scaling is closer to $C_\ell^{\rm GG}\propto\sigma_8^3$.  Thus if one uses cosmic shear to normalize $G(z\ll1)$, then the inferred error on the normalization of the power spectrum is $\sim C_\ell^{\rm GI}/3C_\ell^{\rm GG}$,
\begin{equation}
\Delta\ln G \approx 1.7\frac{b_\kappa}{\Omega_{\rm m}z_{\rm med}^2}.
\end{equation}
The ratio of the contamination in the RSD measurement to that in the CS measurement is then, at low redshifts,
\begin{equation}
\frac{\Delta\ln(fG)|_{\rm RSD}}{\Delta\ln G|_{\rm CS}} \approx
0.6\frac{A}{b_\kappa} \Omega_{\rm m}^{0.4}z_{\rm med}^2.
\label{eq:frac-contam}
\end{equation}
(The factor of 0.6 depends on the shape of the redshift distribution and should be treated only as a rough estimate.)

In the disk galaxy case, we then have from Eq.~(\ref{eq:ABd}) that $0.07A/b_\kappa\approx\psi/K_{\rm a}$, so
\begin{equation}
\frac{\Delta\ln(fG)|_{\rm 
RSD}}{\Delta\ln G|_{\rm CS}} \approx 8\frac\psi{K_{\rm a}}\Omega_{\rm m}^{0.4}z_{\rm med}^2. 
\label{eq:fcd} 
\end{equation}
Thus at low redshift $z\ll 1$ and $\psi/K_{\rm a}\le1$ the RSD measurement suffers less contamination than the CS measurement.  This is because the 
lensing strength for low-redshift surveys is small, and hence the ratio of intrinsic galaxy ellipticities to lensing is enhanced.  However, when the 
source redshift becomes of order unity, Eq.~(\ref{eq:frac-contam}) implies that if $\psi/K_{\rm a}$ is of order unity, then the RSD and CS measurements 
suffer similar levels of contamination.  This statement should be interpreted only in an order-of-magnitude sense, since Eq.~(\ref{eq:fcd}) was derived 
by leaving out factors of $1+z$; more detailed comparisons would require knowledge of how $b_\kappa$, $\psi/K_{\rm a}$, etc. vary with redshift.  
Nevertheless, we can conclude that {\em if we use disk galaxies at $z\sim{\cal O}(1)$, and if the anisotropic selection parameter $\psi/K_{\rm 
a}\sim{\cal O}(1)$, the fractional contamination of the redshift-space distortion measurement due to galaxy alignments is inherently of the same order 
of magnitude as the fractional contamination of the cosmic shear measurement}.

The simple analysis presented here only refers to the uncorrected magnitude of the systematic introduced by tidal alignments.  It is much harder to 
compare the impact of tidal alignments on RSD and CS measurements after corrections are applied, because the nature of the correction is different in 
the two cases.  In CS, intrinsic alignments, even in the nonlinear regime, can be geometrically projected out based on the redshift dependence 
\citep{2004PhRvD..70f3526H}, albeit at the expense of increased error bars and tighter requirements on photometric redshift performance 
\citep{2007NJPh....9..444B}.  In contrast, systematics control for RSD would have to involve a combination of choosing galaxies that are weakly aligned 
($b_\kappa\sim 0$), have selection criteria that are close to isotropic ($\chi,\psi\sim0$), and estimating the residual contamination as well as 
possible.  It would be premature at present to conclude that one of these strategies will be more successful than the other.  Moreover, since CS and 
RSD do not measure the same function (CS measures $G$ whereas RSD measures $fG$) they are not necessarily competitors.

\section{Discussion}
\label{sec:disc}

In this paper, we showed that tidal alignments of galaxies can be a source of systematic error for the redshift-space distortion technique to measure 
the growth of cosmic structures.  This is because a combination of tidal alignments and a viewing-direction-dependent selection function results in a 
different observed power spectrum for Fourier modes in the radial and transverse directions.  We have considered several toy models for the effect.  
The amount of contamination is highly uncertain and survey-dependent; but for surveys using the brightest LRGs for some types of selection criteria the 
effect could be $\sim 10$ per cent, whereas for nearby disk galaxies using the $2\sigma$ upper bounds on tidal alignments and making reasonable 
assumptions about inclination-dependent selection effects we expect $\sim 6$ per cent.  Realistic models would probably give smaller contamination, 
since the tidal alignment effect decreases as one goes down the LRG luminosity function and our result for disk galaxies uses an upper limit. The 
effect is probably unimportant for current RSD measurements, but could represent a problem for future large-scale structure surveys.

There are several strategies to reduce contamination of the RSD by tidal alignments.  One could choose samples of galaxies with weaker alignments; in 
LRG surveys this means going to fainter galaxies, whereas for late-type galaxies we do not have a detection of tidal alignments so no specific 
subsample can be identified as ``best'' at this time.  Secondly, one could try to reduce anisotropic selection effects; for optically thin galaxies 
such as (most) LRGs, this would mean using model magnitudes for selection.  For late-type galaxies whose optical luminosities are significantly 
affected by dust this is harder, but to some extent is possible (e.g. \citealt{2008ApJ...687..976U}).  A final strategy would be to take observed 
density-ellipticity correlation functions and attempt a model-dependent calculation of the $A$ parameter by methods similar to those used in this paper 
(but more sophisticated).

In this paper we have focused our attention on the effect of tidal alignments in the linear regime.  The same processes occur in the quasilinear and 
nonlinear regimes, although there it is even more difficult to make predictions.  We anticipate that future RSD work will make extensive use of the 
quasilinear regime as a test for systematics (e.g. improper finger-of-God compression, second-order biasing, etc.), and possibly to correct ``linear'' 
scales for such effects.  The implications of tidal alignments for quasilinear RSDs are left to future work.

\section*{Acknowledgments}

The author wishes to thank Marc Kamionkowski and Ue-Li Pen for useful feedback; and the referee, Andrew Hamilton, for insightful comments and in
particular asking the question that inspired Appendix~\ref{app:Euler}.

C.H. is supported by the U.S. Department of Energy under contract DE-FG03-02-ER40701, the National Science Foundation under contract AST-0807337, and 
the Alfred P. Sloan Foundation.

\appendix

\section{Reality conditions}
\label{app:real}

This appendix derives the reality condition satisfied by the power spectra in Eq.~(\ref{eq:Pe}).
We denote by $\bar\mR$ the rotation matrix corresponding to $-\bk$ with $-\bar\mR\hat\bk=\ehat_3$ introduced in Section~\ref{ss:stat}, and define ${\mathbfss S}\equiv\bar\mR\mR^{-1}$.

The coefficients in Eq.~(\ref{eq:elm}) are related since $\epsilon(\nhat|\bx)$ is real.
This implies $\tilde\epsilon(\nhat|\bk)=\tilde\epsilon^\ast(\nhat|-\bk)$ and so
\begin{equation}
\sum_{lm} \frac{\rmi^{-l} \tilde\epsilon_{lm}(\bk) Y_{lm}(\mR\nhat)}{\sqrt{2l+1}}
=\sum_{lm} \frac{ \rmi^{l} 
\tilde\epsilon^\ast_{lm}(-\bk) Y^\ast_{lm}(\bar\mR\nhat)}{\sqrt{2l+1}}.
\label{eq:x1}
\end{equation}
Defining $\bomega\equiv\mR\nhat$, we convert Eq.~(\ref{eq:x1}) into
\begin{equation}
\sum_{lm} \frac{\rmi^{-l} \tilde\epsilon_{lm}(\bk) Y_{lm}(\bomega)}{\sqrt{2l+1}}
=\sum_{lm} \frac{ \rmi^{l} 
\tilde\epsilon^\ast_{lm}(-\bk) Y^\ast_{lm}({\mathbfss S}\bomega)}{\sqrt{2l+1}}.
\label{eq:x2}
\end{equation}

By definition $-\bar\mR\hat\bk = \ehat_3$ and $\mR\hat\bk=\ehat_3$, so ${\mathbfss S}\ehat_3=-\ehat_3$.  Since ${\mathbfss S}\in$SO$(3)$ and has a $-1$ eigenvalue, ${\mathbfss S}$ must be a rotation by $\pi$ radians around an axis orthogonal to $\ehat_3$.  We suppose this axis is at position angle $\Phi$, i.e. is $\cos\Phi\ehat_1+\sin\Phi\ehat_2$.  In spherical coordinates, ${\mathbfss S}$ therefore takes the point $\bomega=(\theta,\phi)$ and maps it to the point ${\mathbfss S}\bomega=(\pi-\theta,2\Phi-\phi)$.  We conclude that
\begin{eqnarray}
Y^\ast_{lm}({\mathbfss S}\bomega)
&=& Y_{l,m}^\ast(\phi-\theta,2\Phi-\phi)
\nonumber \\
&=& Y_{l,m}(\phi-\theta,-2\Phi+\phi)
\nonumber \\
&=& (-1)^l Y_{l,m}(\theta,\pi-2\Phi+\phi)
\nonumber \\
&=& (-1)^l \rme^{\rmi m(\pi-2\Phi)} Y_{l,m}(\theta,\phi)
\nonumber \\
&=& (-1)^{l+m} \rme^{-2\rmi m\Phi} Y_{l,m}(\bomega).
\end{eqnarray}
Here the second line used the Condon-Shortley phase convention, the third used the $(-1)^l$ parity of the spherical harmonics, and the fourth used the $\phi$-dependence.  Comparing to Eq.~(\ref{eq:x2}), we obtain
\begin{equation}
\tilde\epsilon_{lm}(\bk) = (-1)^{m} \rme^{-2\rmi m\Phi} \tilde\epsilon^\ast_{lm}(-\bk).
\label{eq:x3}
\end{equation}
Thus the $\bk$ and $-\bk$ components are related, up to a phase $\Phi$ that depends on the choice of $\mR$ and $\bar\mR$.  We may then take the expectation values of products of Fourier modes,
\begin{eqnarray}
\langle \tilde\epsilon^\ast_{lm}(\bk) \tilde\epsilon_{l'm'}(\bk) \rangle &=&
(-1)^{m-m'} \rme^{2\rmi (m-m')\Phi}
\nonumber \\ && \times
\langle \tilde\epsilon^\ast_{lm}(-\bk)\tilde\epsilon_{l'm'}(-\bk') \rangle^\ast.
\end{eqnarray}
Since by rotational invariance the power spectra are only nonzero for $m=m'$, and the power spectra must be the same for $\bk$ and $-\bk$, comparison to Eq.~(\ref{eq:Pe}) shows that the power spectra $P_\epsilon^{ll'm}(k)$ must be real.

\section{General intrinsic alignment power spectrum models}
\label{app:tensor}

This appendix generalizes the results of Sec.~\ref{ss:gps} to intrinsic alignment models with non-scalar contributions ($m\neq 0$ as well as $m=0$) and galaxy models 
lacking inversion symmetry (odd as well as even $l$).

Equation~(\ref{eq:p1}) generalizes to
\begin{eqnarray}
\Pg(\bk) \!\!&=& (b+f\mu^2)^2\Pm(k) 
\nonumber \\ &&
+ 2(b+f\mu^2) \Re \sum_{l=1}^\infty (-\rmi)^l \sqrt{\frac{4\pi}{2l+1}}\,\Pme^l(k) Y_{l0}(\mR\ehat_3)
\nonumber \\ &&
+  \sum_{ll'm} \rmi^{l-l'} \frac{4\pi P_\epsilon^{ll'm}(k)
Y_{lm}^\ast(\mR\ehat_3)Y_{l'm}(\mR\ehat_3)}{\sqrt{(2l+1)(2l'+1)}}.
\label{eq:Tp1}
\end{eqnarray}
The cross-term in Eq.~(\ref{eq:Tp1}) can be simplified by replacing $Y_{l0}$ with a Legendre polynomial using Eq.~(\ref{eq:Lpoly}).
The real part $\Re$ kills the odd-$l$ terms.  Physically the reason these do not contribute to the observed galaxy power spectrum (even if present) is that the 
odd-$l$ intrinsic alignments are $\pi/2$ radians out of phase with the matter perturbations, hence there is no interference in the power spectrum.

The pure intrinsic alignment term in Eq.~(\ref{eq:Tp1}) can also be simplified, albeit with more work.  The $l-l'$ odd terms cancel because $(l,l')$ and $(l',l)$ 
enter with opposite sign (again a manifestation of them being $\pi/2$ radians out of phase).  The $l-l'$ even terms simplify using
\begin{equation}
\frac{4\pi Y_{lm}^\ast(\mR\ehat_3)Y_{l'm}(\mR\ehat_3)}{\sqrt{(2l+1)(2l'+1)}}
= N_{lm}N_{l'm} P_l^m(\mu)P_{l'}^m(\mu),
\end{equation}
where $N_{lm}\equiv \sqrt{(l+m)!/(l-m)!}$.
Thus we finally have
\begin{eqnarray}
\Pg(\bk) \!\!&=& (b+f\mu^2)^2\Pm(k) 
\nonumber \\ &&
+ 2(b+f\mu^2) \sum_{l\ge 2,\rm\,even} (-1)^{l/2} \Pme^l(k) P_l(\mu)
\nonumber \\ &&
+ \sum_{ll',l-l'\rm\,even} (-1)^{(l-l')/2} P_\epsilon^{ll'0}(k) P_l(\mu)P_{l'}(\mu)
\nonumber \\ &&
+  2\sum_{m>0}\sum_{ll',l-l'\rm\,even} (-1)^{(l-l')/2} N_{lm}N_{l'm} P_\epsilon^{ll'm}(k)
\nonumber \\ && \times
P_l^m(\mu)P_{l'}^m(\mu).
\label{eq:Tp2}
\end{eqnarray}
Here we have split off the $m=0$ (scalar) pure intrinsic alignment term, and in the $m\neq 0$ terms we have used the equality of $\pm m$ contributions to restrict the sum to $m>0$ and multiply by 2.

\section{Projected elliptical galaxy}
\label{app:proj}

This appendix derives the projected elliptical galaxy image $I(\bs_\perp)$, Eq.~(\ref{eq:i2}).  The solution presented here is equivalent to that of \citet{1977ApJ...213..368S} in the limit of small 
${\mathbfss W}$ ($t,u\rightarrow 1$ in \citealt{1977ApJ...213..368S} notation). However for clarity we present it here in a different coordinate system and using notation that is well-suited for the 
tidal alignment problem.

We begin by defining a new matrix
\begin{equation}
{\mathbfss U} = \left(\begin{array}{ccc}
\frac12W_{11} & \frac12W_{12} & W_{31} \\ \frac12W_{12} & \frac12W_{22} & W_{32} \\ 0 & 0 & \frac12W_{33}
\end{array}\right).
\end{equation}
Then ${\mathbfss U}+{\mathbfss U}^T={\mathbfss W}$ and to first order in ${\mathbfss W}$,
\begin{equation}
\rho^2 = \bs\cdot({\mathbfss I}+{\mathbfss W})^{-1}\bs
\approx \bs\cdot({\mathbfss I}-{\mathbfss U}-{\mathbfss U}^T)\bs
\approx \bs\cdot({\mathbfss I}-{\mathbfss U})^T({\mathbfss I}-{\mathbfss U})\bs,
\end{equation}
so
\begin{equation}
\rho = \left|({\mathbfss I}-{\mathbfss U})\bs\right|.
\end{equation}
We now define the vector ${\bmath c} \equiv ({\mathbfss I}-{\mathbfss U})\bs_\perp$, which lies in the 12-plane.  Equation~(\ref{eq:i}) simplifies to
\begin{equation}
I(\bs_\perp) = \int {\cal J}\left(\left| {\bmath c} + s_3({\mathbfss I}-{\mathbfss U})\ehat_3 \right|\right) \,\rmd s_3,
\end{equation}
This would be the projected image of the spherical galaxy model (i.e. with ${\mathbfss W}={\mathbfss 0}$) at position ${\bmath c}$ if it were not for the presence of ${\mathbfss U}$.  To first order, the $U_{13}$ and $U_{23}$ terms produce contributions to the integral odd in $s_3$ so they can be dropped, leading to 
\begin{equation}
I(\bs_\perp) = \int {\cal J}\left(\left| {\bmath c} + s_3(1-U_{33})\ehat_3 \right|\right) \,\rmd s_3.
\end{equation}
The factor of $1-U_{33}$ can be eliminate by rescaling the radial variable to $(1-U_{33})s_3$, thereby introducing a factor of the Jacobian $(1-U_{33})^{-1}=1+\frac12W_{33}$.  This leads to Eq.~(\ref{eq:i2}).

\section{Alignment of triaxial galaxy}
\label{app:Euler}

In Sec.~\ref{ss:ellip}, we described the triaxiality of an elliptical galaxy by a traceless-symmetric matrix ${\mathbfss W}$.  It was argued that if there exists a valid Taylor expansion of ${\mathbfss 
W}$ in terms of the local tidal field $s_{ij}$, then one may write:
\begin{equation}
\langle W_{ij}\rangle = 2Bs_{ij} + {\rm higer\ order\ terms},
\end{equation}
i.e. Eq.~(\ref{eq:Wij}).  The purpose of this appendix is to provide an alternative derivation based on the distribution of Euler angles of the galaxy.

Since ${\mathbfss W}$ is a traceless-symmetric matrix, it can be described by its diagonalized form,
\begin{equation}
{\mathbfss W} = {\mathbfss Q}^{\rm T}{\mathbf\Lambda}
{\mathbfss Q},
\end{equation}
where
\begin{equation}
{\mathbf\Lambda} = \left(\begin{array}{ccc}
\lambda_1 & 0 & 0 \\ 0 & \lambda_2 & 0 \\ 0 & 0 & \lambda_3
\end{array}\right)
\end{equation}
is diagonal and
${\mathbfss Q}\in$SO(3)
is an orthogonal matrix specifying the orientation of the galaxy.
Specifically, $Q_{i'i}$ rotates from the ``lab'' (unprimed) frame to a coordinate system aligned with the three principal axes of the galaxy (primed).
It can be expressed in terms of its Euler angles:
\begin{eqnarray}
Q_{11} &=& -\sin\psi\sin\phi+\cos\theta\cos\psi\cos\phi, \nonumber \\
Q_{12} &=& \sin\psi\cos\phi+\cos\theta\cos\psi\sin\phi, \nonumber \\
Q_{13} &=& -\sin\theta\cos\psi, \nonumber \\
Q_{21} &=& -\cos\psi\sin\phi-\cos\theta\sin\psi\cos\phi, \nonumber \\
Q_{22} &=& \cos\psi\cos\phi-\cos\theta\sin\psi\sin\phi, \nonumber \\
Q_{23} &=& \sin\theta\sin\psi, \nonumber \\
Q_{31} &=& \sin\theta\cos\phi, \nonumber \\
Q_{32} &=& \sin\theta\sin\phi, {\rm ~~and} \nonumber \\
Q_{33} &=& \cos\theta.
\label{eq:explicit}
\end{eqnarray}
Tracelessness guarantees $\lambda_1+\lambda_2+\lambda_3=0$ and we can always choose $\lambda_1\le\lambda_2\le\lambda_3$.\footnote{Since
the eigenvectors are only defined up to a sign, there are 4 equivalent choices of ${\mathbfss Q}$ (not $2^3=8$ since we restrict to
proper rotations).  In what follows we assign one of these 4 choices to each galaxy with equal probability.}
An oblate galaxy would have 
$\lambda_1<\lambda_2=\lambda_3$ and a prolate galaxy would have $\lambda_1=\lambda_2<\lambda_3$.  We consider the case of a population of galaxies with fixed 3D axis ratios,
i.e. fixed $\{\lambda_{i'}\}$.  The generalization to a distribution of axis ratios is immediate.

In the absence of a tidal field, the orientation is isotropic, i.e. the probability distribution for the Euler angles is simply the group volume element divided by the volume:
\begin{equation}
\rmd P = \frac{\sin\theta\,\rmd\theta\,\rmd\psi\,\rmd\phi}{8\pi^2},
\end{equation}
where as usual the volume is taken over the range $0\le\theta\le\pi$, $0\le\psi<2\pi$, and $0\le\phi<2\pi$.
In the presence of a tidal field, a general linear-order correction in $s_{ij}$ is:
\begin{equation}
\rmd P = 
[1+s_{ij}C_{ij}(\theta,\phi,\psi)]
\frac{\sin\theta\,\rmd\theta\,\rmd\psi\,\rmd\phi}{8\pi^2}.
\label{eq:PMOD}
\end{equation}
The specific form of the function $C_{ij}$ can be constrained by rotational invariance arguments.  The product $s_{ij}C_{ij}(\theta,\phi,\psi)$ can depend only on the components 
of $s_{ij}$ in the galaxy frame, which are $s_{i'j'}=Q_{i'i}Q_{j'j}s_{ij}$.  That is,
\begin{equation}
\rmd P = 
[1+Q_{i'i}(\theta,\phi,\psi)Q_{j'j}(\theta,\phi,\psi)s_{ij}\bar C_{i'j'}]
\frac{\sin\theta\,\rmd\theta\,\rmd\psi\,\rmd\phi}{8\pi^2}.
\end{equation}
where now all of the dependence on the Euler angles is packaged into ${\mathbfss Q}$, and $\bar C_{i'j'}$ has no dependence on the Euler parameters.
The matrix $\bar C_{i'j'}$ is contracted with the traceless-symmetric tensor $s_{i'j'}$ and hence can be taken to be traceless-symmetric without loss of
generality.  For galaxies with triaxial symmetry, the 180$^\circ$ rotation symmetries around each axis guarantee that $\bar C_{i'j'}$ will be diagonal.\footnote{e.g. 
$\bar C_{12}$ flips sign under 180$^\circ$ rotation around the galaxy 1-axis.}  Hence $\bar{\mathbfss C}$ is traceless-diagonal and has 2 degrees of freedom.  Thus 2 
parameters are required to fully describe the linear alignment of a triaxial object by a tidal field.

(In the special case of an axisymmetric galaxy, e.g. a prolate galaxy with $\lambda_1=\lambda_2$, we have $\bar C_{11}=\bar C_{22}=-\frac12\bar C_{33}$.  In this case, there is only
1 parameter required to describe $\bar{\mathbfss C}$ and hence specify the intrinsic alignment model.)

We next return to the problem of finding the average value of ${\mathbfss W}$:
\begin{eqnarray}
\langle W_{ij} \rangle &=& \Lambda_{i'j'} \langle Q_{i'i}Q_{j'j} \rangle
\nonumber \\ &=& \Lambda_{i'j'}
\int Q_{i'i}Q_{j'j} [1+Q_{k'k}Q_{l'l}s_{kl}\bar C_{k'l'}]
\nonumber \\ && \times
\frac{\sin\theta\,\rmd\theta\,\rmd\psi\,\rmd\phi}{8\pi^2}.
\nonumber \\ &=& \Pi^{(2)}_{iji'j'}\Lambda_{i'j'} + \Pi^{(4)}_{ijkli'j'k'l'}\Lambda_{i'j'}s_{kl}\bar C_{k'l'},
\end{eqnarray}
where we have defined
\begin{equation}
\Pi^{(2)}_{iji'j'} \equiv \int Q_{i'i}Q_{j'j} \frac{\sin\theta\,\rmd\theta\,\rmd\psi\,\rmd\phi}{8\pi^2}
\label{eq:Pi2}
\end{equation}
and similarly for $\Pi^{(4)}_{ijkli'j'k'l'}$.

We next need to evaluate $\Pi^{(2)}_{iji'j'}$ and $\Pi^{(4)}_{ijkli'j'k'l'}$.  Consider first $\Pi^{(2)}_{iji'j'}$.  Since the group volume element is invariant under 
right multiplication, it follows that for any rotation matrix ${\mathbfss O}$ we may replace ${\mathbfss Q}\rightarrow{\mathbfss {QO}}$ in Eq.~(\ref{eq:Pi2}), thus 
establishing:
\begin{equation}
\Pi^{(2)}_{iji'j'} = O_{ia}O_{jb}\Pi^{(2)}_{abi'j'}.
\end{equation}
Now we know that any $3\times 3$ matrix ${\mathbfss X}$ that is rotationally invariant, i.e. ${\mathbfss X}={\mathbfss{OXO}}^{\rm T}$ for all ${\mathbfss O}\in$SO(3),
is proportional to the identity.  For fixed $i'j'$, $\Pi^{(2)}_{iji'j'}$ forms such a $3\times 3$ matrix and hence
$\Pi^{(2)}_{iji'j'} = p_{i'j'}\delta_{ij}$
for some $p_{i'j'}$.  A similar argument using left multiplication shows that $p_{i'j'}$ is proportional to the identity, so
\begin{equation}
\Pi^{(2)}_{iji'j'} = c\delta_{i'j'}\delta_{ij}.
\label{eq:Pi2c}
\end{equation}
Direct integration of Eq.~(\ref{eq:explicit}) for $\Pi^{(2)}_{3333}=\frac13$ enables us to solve for 
$c=\frac13$:
\begin{equation}
\Pi^{(2)}_{iji'j'} = \frac13\delta_{i'j'}\delta_{ij}.
\label{eq:Pi2Result}
\end{equation}

The similar argument for $\Pi^{(4)}_{ijkli'j'k'l'}$ is more complicated.  There are three linearly independent fourth-rank tensor $X_{ijkl}$ that are invariant under 
rotations, i.e. $X_{ijkl} = O_{ia}O_{jb}O_{kc}O_{ld}X_{abcd}$.  A general such tensor can be written as\footnote{This follows from the general theorem, e.g.
Theorem 2.9.A of \citet{Weyl1939}, that all rotation-invariant tensors can be expressed as polynomials in terms of the metric tensor $\delta_{ij}$ and the determinant
tensor $\epsilon_{ijk}$.}
\begin{equation}
X_{ijkl} = \alpha_1 \delta_{ij}\delta_{kl} + \alpha_2 \delta_{ik}\delta_{jl} + \alpha_3 \delta_{il}\delta_{jk}.
\end{equation}
Thus the analogous equation to Eq.~(\ref{eq:Pi2c}) is
\begin{eqnarray}
\Pi^{(4)}_{ijkli'j'k'l'} &=&
 c_1 \delta_{ij}\delta_{kl}\delta_{i'j'}\delta_{k'l'}
+c_2 \delta_{ik}\delta_{jl}\delta_{i'j'}\delta_{k'l'}
\nonumber \\ &&
+c_3 \delta_{il}\delta_{jk}\delta_{i'j'}\delta_{k'l'}
+c_4 \delta_{ij}\delta_{kl}\delta_{i'k'}\delta_{j'l'}
\nonumber \\ &&
+c_5 \delta_{ik}\delta_{jl}\delta_{i'k'}\delta_{j'l'}
+c_6 \delta_{il}\delta_{jk}\delta_{i'k'}\delta_{j'l'}
\nonumber \\ &&
+c_7 \delta_{ij}\delta_{kl}\delta_{i'l'}\delta_{j'k'}
+c_8 \delta_{ik}\delta_{jl}\delta_{i'l'}\delta_{j'k'}
\nonumber \\ &&
+c_9 \delta_{il}\delta_{jk}\delta_{i'l'}\delta_{j'k'}.
\end{eqnarray}
This can be simplified using the permutation symmetries such as $\Pi^{(4)}_{ijkli'j'k'l'} = \Pi^{(4)}_{jiklj'i'k'l'}$.  These imply $c_1=c_5=c_9$ and 
$c_2=c_3=c_4=c_6=c_7=c_8$.  With the constraints $\Pi^{(4)}_{33333333}=\frac15=3c_1+6c_2$ and $\Pi^{(4)}_{11331133}=\frac2{15}=c_1$, we then find $c_1=\frac2{15}$
and $c_2=-\frac1{30}$.

Combined with the tracelessness of ${\bmath\Lambda}$ and ${\mathbfss s}$, diagonality of ${\bmath\Lambda}$, and symmetry of ${\mathbfss s}$, this implies
\begin{equation}
\langle W_{ij} \rangle = \left( \frac15 \sum_{i'=1}^3\bar C_{i'i'}\lambda_{i'}\right) s_{ij}.
\label{eq:WijB}
\end{equation}
The quantity in parentheses is then merely a constant, thus verifying Eq.~(\ref{eq:Wij}).

It is interesting to note that to lowest order in the deviation from sphericity, the viewing direction-dependent selection function will depend only on $\langle 
W_{ij} \rangle$.  Thus the contamination to RSD (and to CS, since to lowest order the intrinsic ellipticity of a galaxy {\em is} $W_{11}-W_{22}+2\rmi W_{12}$) depends 
only on
\begin{equation}
B = \frac1{10} \sum_{i'=1}^3\bar C_{i'i'}\lambda_{i'}.
\end{equation}
Therefore, even though the intrinsic alignments of a triaxial object are described by 2 
parameters (the linearly independent values of $\bar C_{i'i'}$), only 1 linear combination of these is relevant to our analysis.

\end{document}